\documentclass[eqsecnum,amsmath,onecolumn,notitlepage,nofootinbib,superscriptaddress]{revtex4-1}
\usepackage{custom}
\graphicspath{{fig/}}

\begin{document}

\title{Fluctuation dynamics in a relativistic fluid with a critical point}
\author{Xin An}
\email{xan2@uic.edu}
\affiliation{Department of Physics, University of Illinois, Chicago, Illinois 60607, USA}

\author{G\"{o}k\c{c}e Ba\c{s}ar}
\email{gbasar@unc.edu}
\affiliation{Department of Physics and Astronomy, University of North Carolina, Chapel Hill, North Carolina 27599, USA}

\author{Mikhail Stephanov}
\email{misha@uic.edu}
\affiliation{Department of Physics, University of Illinois, Chicago, Illinois 60607, USA}

\author{Ho-Ung Yee}
\email{hyee@uic.edu}
\affiliation{Department of Physics, University of Illinois, Chicago, Illinois 60607, USA}

\date{\today}

\newpage

\begin{abstract}

  To describe dynamics of bulk and fluctuations near the QCD critical
  point we develop general relativistic fluctuation formalism for a
  fluid carrying baryon charge.  Feedback of fluctuations modifies
  hydrodynamic coefficients including bulk viscosity and conductivity
  and introduces nonlocal and non-instantaneous terms in constitutive
  equations. We perform necessary ultraviolet
  (short-distance) renormalization to obtain cutoff independent
  deterministic equations suitable for numerical implementation. We
  use the equations to calculate the universal non-analytic
  small-frequency dependence of transport coefficients due to
  fluctuations (long-time tails). Focusing on the critical mode we
  show how this general formalism matches existing Hydro+ description
  of fluctuations near the QCD critical point and nontrivially extends
  it inside and outside of the critical region.

\end{abstract}

\maketitle

\section{Introduction}
\label{sec:intro}
The classic subject of relativistic hydrodynamics \cite{Landau:2013fluid} has
experienced a renaissance of interest in recent years \cite{Jeon:2015dfa,Romatschke:2017ejr}. 
The interest is driven in large part by the progress in heavy-ion collision
experiments which allow us to create and study droplets of hot and
dense matter governed by physics of strong interaction described by
Quantum Chromodynamics. The increasing body of experimental evidence
that relativistic hydrodynamics is describing the evolution of the
expanding fireball created in these collisions motivates technical
developments as well as a closer look at many fundamental
theoretical concepts in hydrodynamics.

The subject of hydrodynamic fluctuations is particularly relevant to
heavy-ion collisions. The system size~$L$ is not astronomically large
compared to the typical microscopic scale, $\lmic$, (factor 10 at
most is a typical scale separation).\footnote{In the context of
  heavy-ion collisions, $\lmic\sim1/T\sim 1$ fm, while the typical
  hydrodynamic gradient scale is set by the (transverse) size of the
  nucleus $L\sim R\sim 10$ fm.} As a result, fluctuations are large
enough to be easily observable in experiments. In addition, since the
leading corrections to hydrodynamics are due to the nonlinear feedback
of fluctuations, we cannot afford to neglect them -- a luxury one is
used to in ordinary fluid dynamics. Fluctuations are even more
important when enhanced by critical
phenomena.

From the modern point of view, hydrodynamics is a systematic expansion
in spatial gradients. More precisely, it is the expansion of
constitutive equations for stress tensor (and conserved current). The
expansion parameter is the ratio of a typical hydrodynamic wave-number
$k=1/L$ to a microscopic scale, say temperature~$T$, or inverse scattering
length, or, generically,
$1/\lmic$. In this view, the ideal, non-dissipative (i.e., reversible)
hydrodynamics is the truncation of this expansion at lowest (zeroth)
order. At first order in gradients (i.e., at order $k^1$ or, more
precisely, $(k\lmic)^1$) one recovers
standard Landau-Lifshitz or Navier-Stokes hydrodynamics. It is the
following order in this expansion that concerns us here. That order is
not $k^2$, but rather is $k^{3/2}$ (or $k^{d/2}$ in $d$-dimensions).
Such non-analytic in $k$ and, therefore, nonlocal contributions come
from fluctuations in hydrodynamics. Thus it is essential to understand
the physics of hydrodynamic fluctuations to faithfully describe
physics of heavy-ion collisions.%
\footnote{Second order ($k^2$) corrections could be dominant instead
  of fluctuations in special cases, where fluctuations are suppressed,
  as in some large-$N$ theories. Also for dimensions greater than 4,
  fluctuations are parametrically smaller than $k^2$ terms. This paper
  shall be concerned with the generic hydrodynamics in three spatial
  dimensions, relevant for QCD fireball evolution in heavy-ion
  collisions, among other applications.}

Furthermore, in addition to modifying hydrodynamic equations by
effectively nonlocal contributions, the fluctuations themselves are
measured in heavy-ion collision experiments. In particular, one of the
most fundamental questions these experiments aim to answer is the
existence and location of the critical point on the QCD phase
diagram~\cite{Aggarwal:2010cw,Bzdak:2019pkr}. 
The signature of this phenomenon is a certain non-monotonous
behavior of event-by-event fluctuation measures when the parameters of
the collision (such as center of mass energy) is varied in order to
``scan'' QCD phase
diagram~\cite{Stephanov:1998dy,Stephanov:1999zu}. This non-monotonous
behavior is driven by critical phenomena and thus predictable without
being able to determine QCD equation of state at finite density (still
an unsolved theoretical problem).

The existing predictions for critical behavior rely significantly on
the assumption of local thermal equilibrium. However, near the
critical point, the equilibrium is increasingly difficult to achieve
due to critical slowing down and a finiteness of expansion
time. Essentially, this limitation determines the magnitude of the
observable signatures of the critical
point~\cite{Stephanov:1999zu,Berdnikov:1999ph}. Therefore the ability 
to describe the dynamical evolution of fluctuations during
the fireball evolution, in particular, in the proximity the critical
point is crucial. The goal of this paper is to provide such a description.

One of the recent advances towards this goal has been the introduction
of Hydro+ in Ref.~\cite{Stephanov:2018hydro+}, with a recent numerical
implementation in a simplified setup reported in
Ref.~\cite{Rajagopal:2019hydro}. Focusing on the mode
responsible for the critical slowing down, identifying it with the
fluctuation correlator of the slowest hydrodynamic mode, the authors
of Ref.~\cite{Stephanov:2018hydro+} proposed the evolution equation
which describes the relaxation of this non-hydrodynamic
mode to equilibrium. Extending hydrodynamics by addition of such a
mode one is then able to broaden the range of applicability of
hydrodynamics near the critical point and describe the dominant mode
of critical fluctuations at the same time. The crucial ingredient of
this formalism is a nonequilibrium entropy of fluctuations derived in
Ref.~\cite{Stephanov:2018hydro+}.

We approach this problem from a different direction. We start with the
general formalism of relativistic hydrodynamic fluctuations introduced
earlier in Ref.~\cite{An:2019rhf} for neutral (chargeless) fluids and
extend it to include a crucial ingredient -- baryon charge density.
QCD critical point, if it exists, is located at finite
baryon density. The approach we pursue, in which the two-point correlators of
hydrodynamic variables play the role of additional non-hydrodynamic
variables, has been introduced and developed recently in the context of
heavy-ion collisions, but limited to 
special types of flow such as longitudinal boost-invariant
expansion in Refs.~\cite{Akamatsu:2017,Akamatsu:2018,Martinez:2018}. In
a more general but nonrelativistic case this approach was pioneered
by Andreev in the
1970's~\cite{Andreev:1978}. The approach is often referred to as
`hydro-kinetic' to acknowldge the similarity between the two-point
correlators and the distribution functions in kinetic theory. In
particular, the dynamics of the correlators of the pressure
fluctuations is essentially equivalent to the kinetics of the phonon
gas. This physically intuitive picture was the original source of this
formalism~\cite{andreev1970twoliquid,Akamatsu:2017} and was
rigorously derived in general relativistic context in
Ref.~\cite{An:2019rhf}.

The hydro-kinetic approach should be also contrasted with the
traditional stochastic hydrodynamics where the noise is
introduced into hydrodynamic equations \cite{Landau:2013stat2}. From
this point of view, the `hydro-kinetic' approach could also be called
`deterministic', as it replaces stochastic equations with deterministic
equations for the evolution of correlation functions. Of course, the
two approaches solve the same system of stochastic equations, but in
complementary ways. The advantage of the deterministic approach is
that it allows one to deal with the problem of the ``infinite noise'':
the noise amplitude needs to become
 infinitely large as the hydrodynamic cell size is
sent to zero, even though the physical effect of the noise is finite
due to its averaging out in a medium whose properties vary slowly in
space and time. The effect of the infinite (or more precisely cutoff
dependent) noise can be absorbed into renormalization of
hydrodynamic equations -- a procedure which can be performed
analytically in the deterministic approach. This avoids having to deal
with numerical cancellations which would otherwise be necessary in a
direct implementation of stochastic equations.

Near the critical point the deterministic approach we develop here,
although different from Hydro+ in Ref.~\cite{Stephanov:2018hydro+},
nevertheless leads to the description of fluctuations in terms of
two-point correlators as in Hydro+. In this paper we verify that in
the limit of large correlation length the two approaches exactly
match. This is a nontrivial check of the validity of both
approaches. Furthermore, since the deterministic approach is more
general it allows us to extend the Hydro+ approach both closer to the
critical point and further away from the critical point to describe also
ordinary, noncritical fluctuations.

The paper is organized as follows. In
Section~\ref{sec:stoch-hydr-evol} we start from stochastic
hydrodynamics and derive linearized stochastic equations for
fluctuations of hydrodynamic variables. In Section~\ref{sec:kinetic}
we use this result to derive deterministic evolution equations for two-point
correlators of hydrodynamic variables. These equations bear
resemblance to evolution equations in kinetic theory. In
Section~\ref{sec:renormalization} we expand the stochastic hydrodynamic
equations again, now further, to second order in the fluctuations.
Upon averaging over noise, we obtain the equations for averages of
hydrodynamic variables (one-point functions). These equations, due to
nonlinearities, now contain the contributions of two-point
functions. These contributions lead to renormalization of ``bare''
hydrodynamics equations, i.e., they change the ``bare'' equation of
state and ``bare'' transport coefficients into physical quantities. All cutoff
dependence is absorbed at this stage. The remaining contributions are
nonlocal and are known as long-time tails. As an example, we work out explicitly the
non-analytic small-frequency dependence of transport coefficients. 
In Section~\ref{sec:fluctnearcp} we study the behavior of the equations
we derived near the critical point, perform comparison with Hydro+ and
propose an extension to shorter time/length scales which we refer to as
Hydro++.


\section{Stochastic hydrodynamics and fluctuations}
\label{sec:stoch-hydr-evol}

\subsection{Hydrodynamics with noise}
\label{sec:hydr-with-noise}

The starting point of our analysis is the relativistic hydrodynamics
with stochastic noise which drives the thermal fluctuations of
hydrodynamic variables.  The amplitude of the noise is proportional to
the dissipative transport coefficients as a consequence of the
fluctuation-dissipation theorem.  For macroscopically large systems we
are considering, fluctuations are small because they originate from
the noise at microscopically short scales and average out over
macroscopic distances. This allows us to treat the effects of these
fluctuations systematically in a power expansion. We shall discuss the
corresponding power-counting in Section~\ref{sec:conclusions}. In particular, in
this paper we will focus on the quadratic order in fluctuations and
neglect cubic and higher order. In this section, we lay out this
basic set-up, up to the quadratic order of fluctuations that we are
working with. This is a generalization of our previous work for fluids without 
conserved charges \cite{An:2019rhf} to the case of relativistic plasma with a conserved
charge.

Stochastic hydrodynamics with a conserved U(1) charge is defined
by the conservation equations 
\begin{equation}\label{eq:hydro-stoch}
\del_\mu \sT^{\mu\nu}=0,\quad
\del_\mu \sJ^\mu=0.  
\end{equation}
Here, we follow the conventions in
Ref.~\cite{An:2019rhf}, and distinguish stochastic quantities by the
breve accent $\,\breve{}\,$. Defining the stochastic hydrodynamic
variables ($\seps,\sn,\su^\mu$) by the Landau's conditions
\begin{equation}
\label{eq:landau_stoch} 
\sT^{\mu\nu}\su_\nu=-\seps \su^\mu,\quad
\sJ^\mu\su_\mu=-\sn, 
\end{equation}
we can write $\sT^{\mu\nu}$ and $\sJ^\mu$ as
\bea \sT^{\mu\nu}=T^{\mu\nu}(\seps,\sn,\su)+\sS^{\mu\nu},\quad
\sJ^\mu=J^\mu(\seps,\sn,\su)+\sI^\mu,\label{TandJ} \ea in terms of the
``bare" constitutive relations
\begin{eqnarray}\label{eq:constitutive}
{T}^{\mu\nu}(\eps,n,u)&=&\eps \,u^\mu u^\nu +p(\eps,n) \Delta^{\mu\nu}+\Pi^{\mu\nu} \nn
J^\mu(\eps,n,u)&=&n u^\mu+\nu^\mu
\end{eqnarray}
where $ \Delta^{\mu\nu}\equiv g^{\mu\nu}+u^\mu u^\nu$ is the standard
space-like projection operator. The function $p(\eps,n)$ is the
pressure, given by
the equation of state. As usual, $p$, in terms of entropy $s$:
\begin{equation}
  \label{eq:s}
  p = T s - \beps + \mu n,
\end{equation}
where temperature $T$ and chemical potential $\mu$ are defined via
derivatives of $s$ (the
first law of thermodynamics):
\begin{equation}\label{eq:ds}
  d s = \beta d\beps - \alpha dn,\quad
\beta \equiv 1/T, \quad \alpha \equiv \mu/ T\,. 
\end{equation}
Throughout the paper we will also use the enthalpy density,
\begin{equation}
w\equiv\eps+p=Ts+\mu n, \label{eq:w=e+p}
\end{equation}
and, most importantly, entropy per charge ratio
\begin{equation}
  \label{eq:m=s/n}
  m\equiv \frac{s}{n}.
\end{equation}

The constitutive equations in Eqs.~(\ref{eq:constitutive}) 
are organized as an expansion in powers of spatial
gradients. The terms first-order in gradients 
are given by
\begin{eqnarray}\label{eq:viscous}
\Pi^{\mu\nu}&=&-2\eta \left(
\h^{\mu\nu}
-\frac{1}{3} \Delta^{\mu\nu}\divu\right)-\zeta\Delta^{\mu\nu}\divu \\ 
\nu^\mu&=&-\lambda\Delta^{\mu\nu}\partial_\nu\alpha\equiv -\lambda\partial_\perp^\mu \alpha,\label{eq:diffusive-nu}
\end{eqnarray}
and
\begin{equation}
\theta^{\mu\nu}\equiv {1\over 2}\left(\partial^\mu u^\nu+\partial^\nu u^\mu\right),
\quad 
\theta\equiv \partial_\mu u^\mu\equiv\partial\cdot u\,.
\end{equation}
 The shear and bulk viscosities are denoted by $\eta$ and $\zeta$, and
 the charge conductivity is denoted by $\lambda$ \footnote{Note that
   in terms of the conventionally defined conductivity at constant
   temperature defined as $J = -\sigma\del\mu $: $\lambda=\sigma T$.}. The fluctuations are sourced by random noise terms ($\sS^{\mu\nu}$, $\sI^{\mu}$) that are sampled over a Gaussian distribution with an amplitude determined by the fluctuation-dissipation theorem,
\begin{equation}\label{eq:fdt}
\begin{gathered}
\av{\sS^{\mu\nu}(x)}=\av{\sI^\lambda(x)}=0, \quad \av{\sS^{\mu\nu}(x)\sI^\lambda(x^\prime)}=0, \quad \av{\sI^\mu(x)\sI^\nu(x^\prime)}= 2\lambda\bD^{\mu\nu}\delta^{(4)}(x-x^\prime)\,,\\
\av{\sS^{\mu\nu}(x)\sS^{\lambda\kappa}(x^\prime)}= 2T\left[ \eta\, (\bD^{\mu\kappa}\bD^{\nu\lambda}+\bD^{\mu\lambda}\bD^{\nu\kappa})+\left(\zeta-\frac{2}{3} \eta\right) \bD^{\mu\nu}\bD^{\lambda\kappa} \right] \delta^{(4)}(x-x^\prime)\,,
\end{gathered}
\end{equation}
where $\lambda, T, \eta$ and $\zeta$ here assumed be functions of averaged thermodynamic variables, such as energy density and number density. 

Eqs.~(\ref{eq:hydro-stoch}) together with constitutive
equations~(\ref{eq:constitutive}) determine  evolution of
stochastic variables $\seps$, $\sn$ and~$\su$. 
Since we have the freedom to choose an independent 
pair of  scalar variables arbitrarily, we use this
freedom to keep our calculations and resulting equations relatively simple.
We find the following set of variables particularly convenient:
\begin{equation}\label{eq:sm-defined}
\sm\equiv m(\seps,\sn)
\quad\mbox{and}\quad
\breve p \equiv p(\seps,\sn),
\end{equation}
where $m(\beps,n)$ and $p(\beps,n)$ are the entropy per charge and
the pressure, expressed as functions of the energy and charge densities, defined in
Eqs.~(\ref{eq:m=s/n}) and~(\ref{eq:s}). 
This choice simplifies our calculations because the 
fluctuations of $m$ and $p$ are statistically independent in
equilibrium and correspond to two eigenmodes of linearized ideal
hydrodynamic equations.
We shall denote the ensemble averages of these variables by simply
removing the accent, i.e., 
\begin{equation}
  \label{eq:mpu-ave}
  m \equiv \langle \sm \rangle,\quad p \equiv \langle\breve p\rangle,
\quad u  \equiv\langle\su\rangle\,.
\end{equation}
Having defined variables $m$ and $p$ as average values (one-point
functions) of primary variables in Eqs.~(\ref{eq:mpu-ave}) we shall
now define other deterministic variables, which appear in our
equations, such as $\epsilon$ and $n$ as {\em functions} of $m$ and $p$
obtained via equation of state:
\begin{equation}
  \label{eq:emp-nmp}
  \beps \equiv \beps(m,p),\quad n\equiv n(m,p).
\end{equation}
Note that, due to nonlinearities in these relationships,
$\beps\neq \langle\seps\rangle$ and
$n\neq\langle\sn\rangle$.

In order to describe the evolution of these deterministic quantities
we shall perform the ensemble average on the stochastic
equations. Although this eliminates the noise terms, because of the
nonlinearities in the constitutive equations the averaged equations
cannot be simply obtained by substituting stochastic variables by
their averages. We shall describe the effect of these nonlinearities
on the evolution of average values (i.e., one-point functions in
Eq.~(\ref{eq:mpu-ave})) in Section~\ref{sec:renormalization}. These
effects, to lowest order in the magnitude of the fluctuations, are
given in terms of the two-point functions. Our goal in
Section~\ref{sec:kinetic} will be to derive evolution equation for
these correlators. We should also keep in mind that these two-point
functions are of interest in their own right, since they describe the
magnitude of the fluctuations and correlations which, in heavy-ion
collisions, are measurable.

\subsection{Linearized equations}
\label{sec:linearized-equations}

To obtain equations for the two-point functions we first begin by writing
stochastic hydrodynamic equations linearized in deviations of the
stochastic variables from their average values in
Eq.~(\ref{eq:mpu-ave}):
\begin{equation}\label{eq:dm-dp}
\sm=\langle \sm\rangle +\dm,\quad \breve{p}=\langle\breve p\rangle +\dpp,
\quad \su^\mu= \langle \su^\mu\rangle +\du^\mu\,.
\end{equation}
To linear order, the fluctuations of $\sm$ and
$\breve p$ are simply related to fluctuations of
$\seps=\beps(\sm,\breve p)$ and $\sn=n(\sm,\breve p)$ by a
linear transformation with coefficients given by thermodynamic
derivatives. We shall use the following intuitive short-hand notations for these derivatives:
\begin{equation}
  \label{eq:dedp-dmdn}
  d\beps = \beps_m dm + \beps_p dp,\quad
  dn = n_m dm + n_p dp
\end{equation}
whose exact definitions are given in Appendix \ref{sec:derivation}.
Similarly, we find it useful to express the  fluctuations of the 
thermodynamic function
$\breve\alpha=\alpha(\seps,\sn)$  defined in Eq.~(\ref{eq:ds}) in terms of $\delta
m$ and $\delta p$ and define corresponding coefficients:
\begin{equation}
  \label{eq:dalpha-dmdn}
  d \alpha = \alpha_m dm + \alpha_p dp\,.
\end{equation}
Of course, due to nonlinearities in the equation of state the
relationship between fluctuations of  $(\seps,\sn,\breve\alpha)$ and $(\sm,\breve p)$
 is nonlinear, and we
shall deal with this in Section~\ref{sec:renormalization} where we consider the second order terms in the fluctuation 
expansion. 

Now we are ready to expand the constitutive equations to linear order
in fluctuations:
\begin{eqnarray}\label{eq:TJ-dmdp}
\sT^{\mu\nu}&\approx& T^{\mu\nu}(\beps,n,u)+\eps_m\bu^\mu\bu^\nu\dm+\left(\,g^{\mu\nu}+(1+\eps_p)\bu^\mu\bu^\nu\right)\delta p+w\left(\bu^\mu\du^\nu+\bu^\nu\du^\mu\right)
\nn
&&-\eta(\partial_\perp^\mu\du^\nu +\partial_\perp^\nu\du^\mu )-\left(\zeta-\frac{2}{3}\eta\right) \bD^{\mu\nu} \del\cdot \du + \sS^{\mu\nu}\,,
\nn
\sJ^{\mu}&\approx& J^\mu(\beps,n,u)+ n_mu^\mu\dm + n_pu^\mu\dpp +n\du^\mu -\lambda\alpha_m\partial_\perp^\mu\dm-\lambda\alpha_p\partial_\perp^\mu \dpp+\sI^\mu.
\end{eqnarray}
The equations of motion for both the background and the fluctuations
are obtained by substituting Eq.~(\ref{eq:TJ-dmdp}) into
Eq.~(\ref{eq:hydro-stoch}).  By definition, Eq.~(\ref{eq:dm-dp}),
one-point averages of fluctuations vanish,
$\langle \dm\rangle=\langle \delta p\rangle=\langle \delta
u\rangle=0$. Therefore, upon averaging the equations of motion,
$\langle \partial_\mu
\sT^{\mu\nu}\rangle=\langle\partial_\mu\sJ^\mu\rangle=0$, we obtain
\begin{equation}
\label{eq:hydro}
\partial_\mu T^{\mu\nu}(\eps,n,u)=0,\quad 
\partial_\mu J^\mu(\eps,n,u)=0,
\end{equation}
At leading order in gradients, this gives us
equations of ideal hydrodynamics,
\begin{eqnarray}\label{eq:ideal-eom}
	\bu\cdot\partial\eps=-w\divu, \quad \partial_{\perp\mu} p=-wa_\mu, \quad \bu\cdot\partial n=-n\theta,
\end{eqnarray}
which we shall use in the following calculations below. Here $a_\mu\equiv \bu\cdot \del \bu_\mu$ is the fluid acceleration.
Inserting Eqs.~(\ref{eq:hydro}) back into the original
stochastic equations, Eqs.~\eqref{eq:hydro-stoch}, we obtain the linearized equations of motion for
the fluctuations. To present these equations compactly we introduce the relaxation/diffusion coefficients
\begin{equation}
\label{eq:def_gammas}
\begin{aligned}
  \geta\equiv\frac{\eta}{\bw}, \quad \gz\equiv\frac{\zeta}{\bw},\quad
  \gl  
=-\lambda\frac{\alpha_mw}{Tn^2},
\quad \gp=\lambda c_s^2\alpha_p^2Tw\,.
\end{aligned}
\end{equation}
 We also use the thermodynamic relation,
\begin{equation}\label{eq:eos_n/w}
	\frac{w}{n}d\left(\frac{n}{w}\right)=-\frac{1}{w}\left(1-\frac{\alpha_p}{\alpha_m}Tn\right)
        dp-\frac{Tn}{\alpha_mw}d\alpha,
\end{equation}
and express our equations in terms of gradients of $p$ and
$\alpha$.
 With the help of above expressions, we find, after some amount of algebra, the following equations of motion for our fluctuating variables:
\begin{equation}\label{eq:1st_ord_eqs}
\begin{aligned}
  \bu\cdot\partial\dm=&-\frac{1}{\alpha_m}\left(\alpha_p w a_\nu+\partial_{\perp\nu}\alpha\right)\du^\nu +\gl\partial_\perp^{~2}\dm+\frac{\alpha_p}{\alpha_m}\gl\partial_\perp^{~2}\dpp-\frac{1}{Tn}\partial_\mu\bu_\nu\sS^{\mu\nu}+\frac{w}{Tn^2}\partial_\mu\sI^\mu,\\
  \bu\cdot\partial\dpp=&-c_s^2\eps_m\divu(1-\dot{\eps_m})\dm-\left(1+c_s^2+2\dot{c_s}\right)\divu\dpp-w\left[c_s^2\partial_{\perp\nu}-(1-c_s^2)a_\nu\right]\du^\nu \\
  &+\frac{\alpha_m}{\alpha_p}\gp\partial_\perp^{~2}\dm+\gp\partial_\perp^{~2}\dpp-\dot T\partial_\mu\bu_\nu \sS^{\mu\nu}-c_s^2\alpha_pTw\,\partial_\mu\sI^\mu,\\
  \bu\cdot\partial\du_\mu=&-\frac{\eps_m a_\mu}{w} \dm-\frac{1}{w}\left(\partial_{\perp\mu}+\frac{1+c_s^2}{c_s^2}a_\mu+\frac{\partial_{\perp\mu} c_s}{c_s}\right)\dpp-(-\bu_\mu a_\nu+\partial_{\perp\nu}\bu_\mu-c_s^2\Delta_{\mu\nu}\divu)\du^\nu\\ 
  &+\left[\geta \bD_{\mu\nu} \partial_\perp^{~2}+\left(\gz+\frac{1}{3}\geta\right)\partial_{\perp\mu}\partial_{\perp\nu}\right]\du^\nu-\frac{1}{w}\bD_{\mu\nu}\partial_\lambda \sS^{\lambda\nu}\,.
\end{aligned}
\end{equation}
We also introduced a useful notation ``dot'' for the operation defined
as:
\begin{equation}\label{eq:dotX}
 \dot X=\left(\frac{\partial\log X}{\partial\log s}\right)_m=\frac{s}{X}\left(\frac{\partial X}{\partial s}\right)_m
\end{equation}
for a given thermodynamic quantity $X$. Note that since this operation
is a logarithmic derivative it satisfies 
\begin{equation}
{(XY)}\!\dot{\phantom{I}}=\dot X+\dot Y.\label{eq:XYdot}
\end{equation}
This operator appears in our
equations because to leading order (ideal hydrodynamics),
$(u\cdot\partial) \log X = - \dot{X}\theta$ and $\bu\cdot \partial\, m=0$. The quantities $\dot c_s$ and $\dot\eps_m$ involve third order thermodynamic
derivatives (i.e., third derivatives of a $s(\beps,n)$), while the quantity $\dot T$
is a second order thermodynamic derivative (i.e., second derivatives of a $s(\beps,n)$).
Indeed, among all the second order thermodynamic quantities defined in Eqs.~(\ref{eq:dedp-dmdn}),~(\ref{eq:dalpha-dmdn}) and appearing thereafter, only three are independent. In other words, all second order derivatives can be expressed by at most three independent quantities. We find that a convenient choice at this stage of the calculation is $\alpha_m$, $\alpha_p$ and $c_s^2\equiv({\partial p/\partial \eps})_m$. However, at a later stage, we shall instead choose another set, $c_s$, $c_p$ and $\dot T$ (together with independent third order thermodynamic derivatives such as $\dot c_s$, $\dot c_p$), to express our final results. The expressions relating those quantities belonging to different sets are 
\begin{equation}
\begin{gathered}\label{eq:2ndordercoeff}
  \eps_m=(Tn)^2\alpha_p=Tn\left(1-\frac{\dot T}{c_s^2}\right), ~~ \eps_p=c_s^{-2}, ~~ n_m=\frac{(\alpha_pTn-1)Tn^2}{w}=-\frac{\dot T Tn^2}{c_s^2w}, ~~ n_p=\frac{n}{c_s^2w}, ~~ \alpha_m=-\frac{w}{c_pT}.
\end{gathered}
\end{equation}
Using the above relations derived in Appendix \ref{sec:derivation} we can express all coefficients in terms of either chosen set. Although we choose $\alpha_m$, $\alpha_p$ and $c_s^2$ at this stage, in the intermediate steps of the following calculations, we shall sometimes use terms from both sets, if necessary, in order to keep our expressions relatively simple. Choosing $c_s$, $c_p$ and $\dot T$ as the independent quantities will also benefit us when considering particular situations. As discussed in Appendix \ref{sec:comparison}, the meaning of these quantities are more straightforward if one considers a neutral fluid where $\dot T=c_s^2$, or a fluid in the presence of conformal equation of state where $\dot T=c_s^2=1/3$ and $\dot c_s$ vanishes. 

Another commonly known quantity we shall find useful in what follows
is heat conductivity:
\begin{equation}\label{eq:kappa-lambda}
  \kappa=\left(\frac{w}{Tn}\right)^2\lambda
\end{equation}
in terms of which the diffusion coefficient is simply
\begin{equation}
  \label{eq:gl-kappa}
   \gl = \frac{\kappa}{c_p}\,.
\end{equation}

We introduce a collective notation for the fluctuating modes,
\begin{equation}\label{eq:phi_defn}
  \phi_A\equiv(Tn\dm,\dpp/c_s,w\du_\mu),
\end{equation}
where normalization of the modes is chosen to make resulting matrix
equations simpler and more symmetric. We can then write the above
equations for the linearized fluctuations (Eq.~\eqref{eq:1st_ord_eqs})
in a compact matrix form,
\begin{equation}
\bu \cdot \del \dph_A = - \big( \LL+\D+\K \big)_{AB}\dph^B-{\xi}_A \,,
\label{eq:1st_ord}
\end{equation}
where $\LL$, $\D$, and $\K$ are $6\times6$ matrix operators. The operators $\LL$ and $\D$ are the
ideal and dissipative terms, respectively, $\K$ contains the
corrections due to the first-order gradients of background flow, and
six-vector ${\xi}_A$ denotes the random noise. Explicitly
\begin{eqnarray}\label{eq:1st_ord_matrix}
\LL&\equiv &\left( \begin{matrix}
      0 & 0 & 0 \\[1pt]
      0 & 0 & c_s\partial_{\perp\nu} \\[1pt]
      0 & c_s\partial_{\perp\mu} & 0 \\
   \end{matrix}\right)
   ,\quad 
\D\equiv \left( \begin{matrix}
     -\gl\partial_\perp^{~2}~ & ~(c_s\alpha_pTn)^{-1}\gp\partial_\perp^{~2}~ & 0 \\[3pt]
      c_s\alpha_pTn\gl\partial_\perp^{~2} & -\gp\partial_\perp^{~2} & 0 \\[1pt]
     0 & 0 & -\geta \bD_{\mu\nu} \partial_\perp^{~2} -( \gz+\frac{1}{3}\geta)\partial_{\perp\mu}\partial_{\perp\nu}\\
   \end{matrix}\right),   
   \nn
   \nn
\K&\equiv &
\left( \begin{matrix}
 (1+\dot T)\divu & 0 & \frac{Tn}{\alpha_m}\left(\alpha_p a_\nu+\frac{1}{w}\partial_{\perp\nu}\alpha\right) \\[4pt]
 c_s\alpha_p(1-\dot{\eps_m})Tn\divu & (1+c_s^2+\dot{c_s})\bdivu & \left(2-\frac{(\alpha_pTn)^2}{\alpha_m}\right)c_sa_\nu-\frac{c_s\alpha_pT^2n^2}{\alpha_m w}\partial_{\perp\nu}\alpha  \\[2pt]
 \alpha_pTn a_\mu &  \frac{1+c_s^2}{c_s}a_\mu+\partial_{\perp\mu} c_s  &  -\bu_\mu a_\nu +\partial_{\perp\nu}\bu_\mu+\bD_{\mu\nu}\bdivu\\
   \end{matrix}\right),
   \nn
    \nn
   {\xi}&\equiv & \left(-\frac{w}{n}\partial_\lambda\sI^\lambda,\, c_s\alpha_pTw\partial_\lambda\sI^\lambda,\, \bD_{\mu\kappa}\del_\lambda  \sS^{\lambda\kappa}\right).
\end{eqnarray}
Equation (\ref{eq:1st_ord}) for linearized fluctuations provides the foundation for the fluctuation evolution equations for the two-point correlation functions, derived in the next section.


\section{Fluctuation kinetic equations}
\label{sec:kinetic}
The physical effects of fluctuations on hydrodynamic flow manifest themselves through two-point functions. This is because, by definition, the first order fluctuations average to zero (i.e. $\av{\phi_A(x)}=0$ via Eq.~\eqref{eq:dm-dp}) and the leading order corrections to $\av{\sT^{\mu\nu}}$ and $\av{\sJ^\mu}$ come from the second order terms in the fluctuation expansion (i.e. the two-point functions) whose time evolution equation we derive in this section. How these two-point functions modify the hydrodynamic flow, in other words the feedback of fluctuations on background flow, will be discussed in Section~\ref{sec:renormalization}.

Our strategy is to use equations of motion for linearized
fluctuations, Eq.~\eqref{eq:1st_ord}, to derive an evolution equation
for the ``equal-time"  two-point correlation function of
fluctuations,  $\av{ \phi_A(x^+)\phi_B(x^-) }$, obtained by averaging
over the statistical ensemble generated by the stochastic
noises. Before we do so, we discuss some general features of  the
two-point correlator of hydrodynamic variables
\begin{equation}\label{eq:GAB-def}
G_{AB}(x,y)\equiv \av{ \phi_A(x^+)\phi_B(x^-) }, \quad
\mbox{where} \quad x^\pm = x\pm y/2\,.
\end{equation}
In a static homogeneous equilibrium state of the fluid, the correlator
is translationally invariant, i.e., depends only on the separation
$y= x^+-x^-$ and not on the midpoint position
$x=(x^++x^-)/2$. Furthermore, because equilibrium correlation length
is shorter than the coarse grained resolution of hydrodynamics, the
equilibrium equal-time correlation function is essentially a delta
function of the separation vector $y$ with the magnitude determined by
the the well-known functions of average thermodynamic variables
$\beps$ and $n$. 

However, in a generic {\em relativistic} hydrodynamic flow several new
observations need to be made. First of all, the concept of ``equal
time'' is no longer obvious, as it depends on the frame of
reference. The most natural choice, the rest frame of the fluid, is
now locally different in different points. We will discuss how to
implement it in the next subsection, following the formalism
introduced in Ref.~\cite{An:2019rhf}.  

Furthermore, not only the local thermodynamic conditions, and thus
{\em equilibrium} magnitude of fluctuations, slowly vary in
space-time, but also the fluctuations themselves are driven {\em out}
of equilibrium. Therefore, not only the fluctuation correlator depends
slowly on $x$, but it also acquires nontrivial $y$ dependence, beyond
the equilibrium delta function. It is crucial that the scale of that
$y$ dependence is short compared to the scale of the dependence on
$x$. 

The estimate of the $y$-dependence scale can be made by observing that
the equilibration of fluctuations of hydrodynamic variables is a
diffusive process (since the variables obey conservation equations).
This means that the scale of
equilibration\footnote{Ref.~\cite{An:2019rhf} uses notation
  $\ell_{\rm eq}$ for this scale. $\ell_*\equiv\ell_{\rm eq}$.}
$\ell_*$ is the diffusion length during time interval characteristic
of the evolution. For the reciprocal quantities such as fluctuation
wave-number $q_*\equiv1/\ell_*$ and the frequency $c_s k$ of the
sound, one obtains $\gamma q_*^2\sim c_s k$ and thus
$q_*=\sqrt{c_s k/\gamma}\gg k$. In other words, $\ell_*\ll L \equiv 1/k$. This
separation of scales of $y$ and $x$ dependence of the correlation
function, or between characteristic wave-numbers $q$ of the
fluctuations and $k$ of the background will be used to systematically organize our calculations and
results in the form of an expansion in $k/q\ll 1$ as well as $k\lmic\ll1$. Note that, for the
characteristic wave-numbers of the fluctuations and the background, the
ratio $k/q\sim (k\lmic)^{1/2}$. In other words, this expansion is controlled by
a power of the same small parameter as the hydrodynamic gradient
expansion itself.

With this separation of scales in mind, it is convenient to work with
the Wigner transform of $G_{AB}(x,y)$, that is essentially the Fourier
transform with respect to (spatial components of) $y$, which we shall
label as $W_{AB}(x,q)$.  Since $q$ corresponds to the wave vector of
fluctuating modes that contribute to $G_{AB}$, it is similar in
concept to the momentum of a particle in quantum mechanics. In this
quantum mechanical analogy, the Wigner
transform would be the (matrix valued) phase space distribution
of the fluctuation modes or a density matrix in phase space, $(x,q)$,
in an effective kinetic theory of fluctuation quanta. The evolution equation
of $W_{AB}(x,q)$, which is derived in this section, closely resembles
a Boltzmann-type kinetic equation for the fluctuation degrees of
freedom that are, in the case of hydrodynamics, phonons. In our
previous work, Ref.~\cite{An:2019rhf}, we derived and studied such an
equation for relativistic hydrodynamics {\em without a conserved
  charge}. In this paper we present its generalization to the case
with a conserved U(1) charge, which contains additional nontrivial
features we will discuss in the subsequent sections
\ref{sec:renormalization} and \ref{sec:fluctnearcp}. Some of these
features, such as the existence of the slow scalar mode, play an
important role in the critical dynamics near the QCD critical point
which we discuss in Section~\ref{sec:fluctnearcp}.  The derivation in
this section closely parallels the analysis presented in
Ref.~\cite{An:2019rhf}. For completeness of this paper we will briefly
summarize the key concepts and steps from Ref.~\cite{An:2019rhf} here
before presenting our final result at the end.

\subsection{Confluent correlator and confluent derivative}
\label{sec:covar-deriv-conn}

The concepts of ``equal-time'' and ``spatial" $y$ coordinates we
invoke when defining $G_{AB}(x,y)$ and its Wigner transformation in
the above discussion become nontrivial in a general background
of relativistic flow. Both concepts require choosing a frame of
reference. The most natural choice -- the local rest frame of the
fluid, characterized by the (average) fluid velocity~$u^\mu(x)$ --
varies point to point with $x$. The change of the frame from point to
point is responsible for changing the values of various vector
components of hydrodynamic fluctuations $\phi_A$, such as
$\delta u_\mu$, entering in the definition of $G_{AB}$ in
Eq.~(\ref{eq:GAB-def}). This variation is purely kinematic (Lorentz
boost) and has nothing to do with the local dynamics of fluctuations
that we are interested in. Our goal it to define a measure of
fluctuations and a measure of its changes with space and time to be
independent of such mundane kinematic effects. We achieve this by
introducing the notions of ``confluent correlator'' and ``confluent
derivative" which we describe below, summarizing
Ref.~\cite{An:2019rhf}.

The key to defining these new concepts is a parallel transport or,
equivalently a connection, that takes care of the change of $u(x)$
between two points, say, $x$ and $x+\Delta x$. We introduce a boost
$\Lambda(\Delta x)$ which maps $u(x+\Delta x)$ to
$u(x)$, i.e.,
\begin{equation}
\Lambda(\Delta x) u(x+\Delta x)=u(x)\,.\label{eq:Lambda-u=u}
\end{equation}
In principle, this
boost is not unique.  In our previous work, Ref.~\cite{An:2019rhf}, we propose to use the most natural choice, that is
a pure boost without a spatial rotation in the local rest frame of
$u(x)$. 
Note that the boost in Eq.~(\ref{eq:Lambda-u=u}) is defined
for arbitrary $\Delta x$. In practice, however, we only need its
infinitesimal form:
\begin{equation}\label{eq:Lambda-boost}
\Lambda^{\phantom\mu\nu}_\mu(\Delta x)=\delta^{\phantom \mu\nu}_\mu-u_\mu\Delta u^\nu+\Delta u_\mu u^\nu,\quad \Delta u^\mu\equiv u^\mu(x+\Delta x)-u^\mu(x)=\Delta x^\alpha(\partial_\alpha u^\mu).
\end{equation}

Next, we introduce the notion of ``confluent
correlator''. This notion arises because a certain property of the ``raw''
definition of the two-point correlator $G_{AB}$  prevents us from
cleanly separating $x$ and $y$ dependence and performing Wigner
transform. Specifically 
\begin{equation} 
u^A(x^+)G_{AB}(x,y)=G_{AB}(x,y)u^B(x^-)=0,
\label{eq:orthogonality}
\end{equation}
where $u_A\equiv(0,0,u_\mu)$. These constraints follow from the
orthogonality $u^\mu(x^\pm)\du_\mu(x^\pm)=0$ and relate different
vector components of $G_{AB}(x,y)$ in a $y$-dependent way.
We can deal with this problem by using Eq.~(\ref{eq:Lambda-u=u}) to boost
$G_{AB}(x,y)$ in such a way that instead of being orthogonal to
$u^A(x^+)$ and $u^B(x^-)$, it is orthogonal to $u^A(x)$ and $u^B(x)$.
Thus we define the ``confluent correlator'' as
\begin{equation}
  \label{eq:LGL}
  \GG_{AB}(x,y) \equiv\,\Lambda^{\phantom AC}_A(y/2)\,\Lambda^{\phantom BD}_B(-y/2)\, \N_{CD}(x,y),
\end{equation}
where $\Lambda^{\phantom AC}_A(\Delta x)=\Lambda^{\phantom
  \mu\nu}_\mu(\Delta x)$ when $AC=\mu\nu$, and an identity
transformation otherwise.
It is straightforward to check that the confluent correlator indeed satisfies
\begin{equation}\label{eq:uG=Gu=0}
u^A(x)\GG_{AB}(x,y)=\GG_{AB}(x,y)u^B(x)=0,
\end{equation}
i.e., the constraints are now independent of $y$. This allows us to
meaningfully perform the Wigner transformation of this object with
respect to $y$ coordinates without affecting the
constraint. Correspondingly, the confluent Wigner function, defined as
a Fourier transform on the locally spatial hyper-surface
$u(x)\cdot y=0$, is given by
\begin{equation}
  \label{eq:Wigner-function}
  \W_{AB}(x,\p) \equiv \int d^4y
 \,\delta( u(x)\cdot y)\, e^{-i\p\cdot y}
\, \GG_{AB}(x,y)\,,
\end{equation}
and obeys
\begin{equation}
  \label{eq:uW=Wu=0}
  u^A(x)\W_{AB}(x,q)=\W_{AB}(x,q)u^B(x)=0.
\end{equation}

Note that, although the wave vector $q$ is a four-vector, $\W_{AB}$
depends only on its projection on the hyper-plane defined by $u(x)\cdot
q=0$.  
In order to eliminate the redundant component along $u(x)$ we
can impose the constraint $u(x)\cdot q=0$.  Because this
constraint depends on $u(x)$, a meaningful derivative of $W_{AB}(x,q)$
with respect to $x$ should be then defined with a parallel transport of $q$
by $\Lambda(\Delta x)$ from Eq.~(\ref{eq:Lambda-u=u}) to maintain the
constraint. We shall also use the same transport to eliminate the purely
kinematic effect of the boost on the vector components of variables $\phi_A$. This leads to the notion of ``confluent
derivative", $\cfd_\mu W_{AB}(x,q)$, which we define as 
\begin{equation} 
\Delta
x^\mu\cfd_\mu W_{AB}(x,q)\equiv \Lambda(\Delta x)^{\phantom AC}_A
\Lambda(\Delta x)^{\phantom BD}_B W_{CD}(x+\Delta x,\Lambda(\Delta
x)^{-1} q)-W_{AB}(x,q).
\label{wignerconf}
\end{equation}
It is straightforward to see that $\cfd_\mu W_{AB}(x,q)$ is equal to the Wigner transformation of the confluent derivative $\cfd_\mu\GG_{AB}(x,y)$ similarly defined by
\be
\Delta x^\mu\cfd_\mu G_{AB}(x,y)\equiv \Lambda(\Delta x)^{\phantom AC}_A \Lambda(\Delta x)^{\phantom BD}_B \GG_{CD}(x+\Delta x,\Lambda(\Delta x)^{-1} y)-\GG_{AB}(x,y).
\ee

Although the above confluent derivative is well defined conceptually,
its practical evaluation requires us to introduce a local basis in the
spatial hyper-surface of $u(x)\cdot q=0$ at each point $x$, the triad
$e^\mu_a(x)$ ($a=1,2,3$) satisfying $u(x)\cdot e_a(x)=0$ and
$e_a\cdot e_b=\delta_{ab}$. The choice of the triad field $e_a(x)$ is
arbitrary and different choices are related by local SO(3)
rotations. Using this basis, we can write $q=e_a(x) q^a$ with an
internal three-vector ${\bm q}=\{q^a\}\in {\mathbb R}^3$, and consider
$W(x,q)$ as a function of $\bm q$:
\begin{equation}
W(x,\bm q)\equiv W(x,q=e_a(x) q^a).\label{eq:W-bmq}
\end{equation}

Working out (\ref{wignerconf})
explicitly, we obtain
\begin{equation}
  \label{eq:nabla-mu-wigner}
  \cfd_\mu \W_{AB}(x,\bm q) = \partial_\mu  \W_{AB} 
- \ucon_{\mu A}^C \W_{CB} - \ucon_{\mu B}^C \W_{AC}
+ \econ_{\mu a}^{b}\, \p_b \frac{\partial}{\partial \p_a} \W_{AB},
\end{equation}
where the connection, \be \ucon^\nu_{\lambda\mu} \equiv
u_\mu\partial_\lambda u^\nu -u^\nu\partial_\lambda u_\mu, \ee arises
from the Lorentz boost acting on indices $A$ and $B$, whereas the internal SO(3)
connection, 
\begin{equation}
\econ^b_{\lambda a} \equiv e^b_\mu \partial_\lambda
e^\mu_a,
\end{equation}
is due to the fact that $\Lambda(\Delta x)^{-1}e_a(x)$ is
in general not equal to $e_a(x+\Delta x)$, but could be related by an
additional SO(3) rotation.\footnote{One can introduce
  $W_{ab}(x,\bm q)$ by $W_{AB}=e_A^a e_B^b W_{ab}$, so that the
  confluent derivative of $W_{ab}$ involves only the SO(3) connection
  $\econ$. In other words, $\ucon$ reduces to a SO(3) connection when
  it acts on $W_{AB}$. It is a simple matter of choice to work with
  $W_{ab}(x,\bm q)$ or with $W_{AB}$.} Note that the partial
derivative $\partial_\mu$ in (\ref{eq:nabla-mu-wigner}) is taken at
fixed $\bm q$, {\em not} fixed $q=e_a(x)q^a$.  From here on we will
 use the notation $W_{AB}(x,q)$, with understanding that it is a
 function of $x$ and $\bm q$ given by 
 Eq.~(\ref{eq:W-bmq}).

\subsection{Kinetic equation for the Wigner function $W_{AB}$}
\label{sec:kin-eq}
Having introduced the necessary mathematical tools, we now derive the
evolution equation of $W_{AB}(x,q)$ by using, as the starting point,
the stochastic equation of motion for linearized fluctuations,
Eq.~(\ref{eq:1st_ord}). The crux of this derivation is expressing the 
equation for $W_{AB}$ in terms of the confluent derivatives, which
have a clear physical meaning as the derivatives in the co-moving
frame. This leads to many nontrivial cancellations.

We start with the evolution equation for
the two-point function $G_{AB}(x,y) = \langle\dph_A(x^+))\dph_B(x^-)\rangle$,
with $x^\pm=x\pm y/2$, and choose $y$ to be spatial in the frame $u(x)$.
The time evolution of $\N_{AB}(x,y)$ is obtained by
\begin{equation}
u(x)\cdot\del \N_{AB}(x,y)=\av{(u(x)\cdot\del\dph_A(x^+))\dph_B(x^-)}+\av{\dph_A(x^+)(u(x)\cdot\del\dph_B(x^-))},\label{naiveGeq}
\end{equation}
where the derivative operator,
$\del$, always acts on the {\em first} argument of the function,
such as $x$ in $G(x,y)$, or $x^\pm$ in $\phi(x^\pm)$. 
Derivative with respect to the second
argument, if there is any, will be labeled explicitly. Next, we convert the time
derivatives in the right hand side of (\ref{naiveGeq}) into spatial
derivatives. In order to do so we have to expand $u(x)=u(x^\pm)\mp
{1\over 2} y\cdot\partial
u(x)$ and use the evolution equation for the one point function,
Eq. \eqref{eq:1st_ord}.  To perform the resulting averaging in the r.h.s. of
Eq.~(\ref{naiveGeq}) we need to know the average of the two-point
function of the noise, which can be calculated using the definition in
Eq.~\eqref{eq:1st_ord_matrix} and Eq.~\eqref{eq:fdt}:
\begin{equation}
  \av{\xi_A(x^{\prime})\xi_B(x^{\prime\prime})}=2\Q_{AB}\delta^{(4)}(x^{\prime}-x^{\prime\prime}),
\end{equation}
where
\begin{equation}\label{FDrel}
    \Q =Tw\left( \begin{matrix}
    \alpha_m^{-1}\gl\partial_\perp^{~2} & (c_s\alpha_pTn)^{-1}\gp\partial_\perp^{~2} & 0 \\[4pt] 
    (c_s\alpha_pTn)^{-1}\gp\partial_\perp^{~2} & -\gp\partial_\perp^{~2} & 0 \\[3pt]
    0 & 0 & -\left(\gamma_\eta\bD_{\mu\nu}\partial_\perp^{~2}+\left(\gamma_\zeta+\frac{1}{3}\gamma_\eta\right)\partial_{\perp\mu}\partial_{\perp\nu}\right) \\
   \end{matrix}\right).
\end{equation}
Proceeding from Eq.~(\ref{naiveGeq}) along these steps we arrive at
\begin{eqnarray}
\label{eq:kinetic_eq_y}
\bu \cdot \del \N_{AB}(x,y) &=& -\big(\LL^{(y)}+\frac{1}{2}\LL+\D^{(y)}+\K+\Y\big)_{AC} \N^C_{\,\,\,B}(x,y) -  \big(-\LL^{(y)}+\frac{1}{2}\LL+\D^{(y)}+\K+\Y\big)_{BC} \N_A^{\,\,\,C}(x,y) \nn
&&+\lim_{\delta t\to0}\frac{1}{\delta t}\int_{\bu\cdot x^+}^{\bu\cdot x^+ +\delta t}  \bu\cdot dx^\prime \int_{\bu\cdot x^-}^{\bu\cdot x^-+\delta t}  \bu\cdot dx^{\prime\prime} \av{\xi_A(x^{\prime})\xi_B(x^{\prime\prime})} \nn
&=& -(\LL^{(y)}+\frac{1}{2} \LL +\D^{(y)}+\K+\Y)_{AC}\, \N^C_{\,\,\,B}(x,y) -(-\LL^{(y)}+\frac{1}{2} \LL +\D^{(y)}+\K+\Y)_{BC}\N_A^{\,\,\,C}(x,y)\nn
&& +2 \Q^{(y)}_{AB} \delta^3(\yp),\qquad\,\
\end{eqnarray}
where the superscript
 $(y)$ on an operator indicates that the derivatives within that
 operator act on $y$, the second argument of $G_{AB}(x,y)$. For example,
\begin{equation}
\label{eq:L_y}
\LL^{(y)}\equiv\left( \begin{matrix}
      0 & 0 & 0\\
      0 & 0 & c_s(x)\partial_{\perp\nu}^{(y)} \\
     0 & c_s(x)\partial_{\perp\mu}^{(y)} & 0 \\
   \end{matrix}\right).
\end{equation}
The matrix $\Y$,
\begin{equation}
\Y\equiv \left( \begin{matrix}
    \bD_{\lambda\kappa} & 0 & 0 \\
    0 & (1-c_s^2)\bD_{\lambda\kappa} & c_s\bu_\nu\bD_{\lambda\kappa} \\
    0 & c_s\bu_\mu\bD_{\lambda\kappa} & \bD_{\mu\nu}\bD_{\lambda\kappa}-c_s^2\bD_{\mu\lambda}\bD_{\nu\kappa} \\
   \end{matrix}\right)\frac{1}{2}y\cdot\partial\bu^\lambda\partial_\perp^{(y)\kappa} +\frac{1}{2c_s}y\cdot\partial_\perp c_s\LL^{(y)},
\end{equation}
results from the $y$-dependence in $\bu(x^\pm)$ and $c_s(x^\pm)$. Note
that in deriving Eq. (\ref{eq:kinetic_eq_y}), we neglected higher
order terms in $y$, based on the scale separation between background
wave-number $k$ and fluctuation wave-number $q$: $(\partial u)y\sim (\partial c_s)y\sim k/q \ll 1$. 

Eq.~(\ref{eq:kinetic_eq_y}) for $G_{AB}$ can then be used to derive
the evolution equation for the Wigner function $W_{AB}$, while expressing all derivatives in terms of the confluent
derivatives. After some algebraic manipulations, we find the following
result, as expressed in matrix form, 
\begin{equation}
\begin{aligned}\label{eq:kinetic_eq}
\bu \cdot \cfd \W(x,\p)&= -\comm{ i \LL^{(\p)}, \W}
-  \mcomm{\frac{1}{2}\Lcon+\D^{(\p)}+\K', \W}
+2 \Q^{(\p)}+(\partial_{\perp\lambda} \bu_\mu) \pp^\mu \frac{\del \W}{\del \pp_{ \lambda}} \\
& +\frac{1}{2}\left(a_\lambda+\frac{\partial_{\perp\lambda}c_s}{c_s}\right) \acomm{\LL^{(\p)},  \frac{\del \W}{\del \pp_{ \lambda}}}
+\frac{\del}{\del \pp_{ \lambda}} \left(\mcomm{ \Om_\lambda,\W}-\frac{1}{4} [\Hmatrix_\lambda,[\LL^{(\p)},\W]] \right),\quad
\end{aligned}
\end{equation}
where $[A,B]=AB-BA$ and $\{A,B\}=AB+BA$ are the usual matrix (anti)
commutators, while a new notation is introduced for the
anti-commutator which appears naturally in this context:
\begin{equation}
  \label{eq:mixed-comm}
\mcomm{A,B} \equiv AB + BA^\dag 
\,.
\end{equation}
The matrices that appear in Eq.~\eqref{eq:kinetic_eq} read
\begin{eqnarray}
&&\LL^{(\p)}\equiv c_s\left( \begin{matrix}
    0 & 0 & 0 \\
    0 & 0 &  \pp_{\nu} \\
    0 &\pp_{\mu} & 0 \\
   \end{matrix}\right)
   ,\quad 
\Lcon^{(x)} \equiv  c_s\left( \begin{matrix}
    0 & 0 & 0 \\
    0 & 0 &  \cfd_{\perp\nu} \\
    0 & \cfd_{\perp\mu} & 0 \\
   \end{matrix}\right)
   ,\quad \nn\nn
&&\D^{(\p)}\equiv \left( \begin{matrix}
      \gl q^2 & -(c_sTn\alpha_p)^{-1} \gp q^2 & 0 \\[3pt]
     - c_sTn\alpha_p \gl q^2 & \gp q^2 & 0 \\
     0 & 0 & \geta \Delta_{\mu\nu} \pp^2 +\left( \gz+\frac{1}{3}\geta\right)\pp_{\mu} \pp_{\nu} \\
   \end{matrix}\right),   
    \nn\nn
&&\Q^{(q)}\equiv Tw\left( \begin{matrix}
    -\alpha_m^{-1}\gl q^2 & -(c_s\alpha_pTn)^{-1}\gp q^2 & 0 \\[4pt] 
    -(c_s\alpha_pTn)^{-1}\gp q^2 & \gp q^2 & 0 \\[3pt]
    0 & 0 & \left(\gamma_\eta\bD_{\mu\nu}q^2+\left(\gamma_\zeta+\frac{1}{3}\gamma_\eta\right)q_\mu q_\nu\right) \\
   \end{matrix}\right),
   \nn\nn
&&\K'\equiv\K+\Delta\K, \quad \Delta\K \equiv -\frac{\theta}{2}\mathbb{1}
-\frac{1}{2}\left( \begin{matrix}
0 & 0 & 0 \\[5pt]
0 & 0 &  c_sa_\nu+\partial_{\perp\nu}c_s \\[3pt]
0 &  c_sa_\mu+\partial_{\perp\mu}c_s & - 2u_\mu a_\nu\\
   \end{matrix}\right),
\nn\nn
&&\Om_\lambda\equiv c_s^2\left( \begin{matrix}
   0 & 0 & 0 \\
    0 &  \omega_{\kappa\lambda}\pp^\kappa & 0 \\
    0 & 0 & \omega_{\mu\lambda} \pp_\nu  \\
   \end{matrix}\right),\quad
 \Hmatrix_\lambda\equiv c_s\left( \begin{matrix}
      0 & 0 & 0 \\
      0 & 0 &  \partial_\nu\bu_\lambda \\
    0 & \partial_\mu\bu_\lambda & 0 \\
   \end{matrix}\right) ,
   \label{eq:matrices}
\end{eqnarray}
where $\K$ is defined in Eq.~\eqref{eq:1st_ord_matrix} and $\omega_{\mu\nu}$ is the fluid vorticity,
\begin{equation}\label{eq:omega-def}
\omega_{\mu\nu}=\frac{1}{2} ( \del_{\perp \mu} \bu_\nu-
\del_{\perp\nu} \bu_\mu).
\end{equation}

Equation~(\ref{eq:kinetic_eq}) is linear in $W$. The inhomogeneous
term $2\Q^{(q)}$ is the source for random noise, and the matrix
$\D^{(q)}$ characterizes dissipation. In a static uniform background,
the balance between the two gives the equation for the 
equilibrium value for the Wigner function:
\begin{equation}
 -\comm{ i \LL^{(\p)}, \W^{(0)}} - \left[\D^{(\p)},\W^{(0)}\right\} + 2\Q^{(\p)} =0\label{eq:QDW}\,.
\end{equation}
which is a fluctuation-dissipation relation.
This equation is solved by
\begin{equation} 
  W^{(0)}=T\bw\left( \begin{matrix}
      {c_pT}/{w} & 0 & 0 \\[2pt]
    0 & 1 & 0 \\[1pt]
    0 & 0 & \bD_{\mu\nu} \\
\end{matrix}\right) ,\label{eqvalue}
\end{equation}
where we used Eq.~\eqref{eq:2ndordercoeff}: $\alpha_m^{-1} = - c_pT/w
$. Eq.~(\ref{eqvalue}), taken together with Eq.~(\ref{eq:phi_defn}), is in agreement with the well-known thermodynamic expectation values: $V\av{(\delta m)^2}=c_p/n^2$, $V\av{(\delta
  p)^2}=c_s^2Tw$, $V\av{(\delta u)^2}=T/w$ and $\av{\delta m\delta p}=\av{\delta
  m\delta u}=\av{\delta p \delta u}=0$, where $V$ is the volume of the system.
The matrices $\K'$, $\Om_\lambda$, and
$\Hmatrix_\lambda$ encode the effects of background gradients, that
drive the system out of equilibrium.  

\subsection{Averaging out fast modes}
\label{sec:diagonalization}

Some of the components of $W_{AB}(x,q)$ oscillate fast with a
characteristic frequency $\omega\sim c_s q$, due to the
$\LL^{(\p)}\sim c_s q$ term in the matrix kinetic equation
(\ref{eq:kinetic_eq}). According to our hierarchy of scales, the other
terms in (\ref{eq:kinetic_eq}), are of order either $k$ or
$\gamma \p^2$ which are smaller than this oscillation frequency,
$c_s q$. This separation of time scales leads to a new effective
description of the system where the fast components of $W_{AB}$ are
eliminated by time averaging and only slow modes remain. The
corresponding coarse-graining time scale $\ab_t$
satisfies
\begin{equation}
  \label{eq:at}
  c_s k \ll \ab_t^{-1} \ll c_s \p.
\end{equation}
The slow components of $W_{AB}$ that survive time averaging correspond to effective distribution functions in a Boltzmann-like kinetic theory of fluctuations.
Note that this is also similar to how we diagonalize a quantum density
matrix to identify the particle distribution functions starting from quantum field theory.

To identify the fast components, we express the kinetic equation in the basis where $\LL^{(\p)}$ is diagonal. $\LL^{(\p)}$ has six eigenvalues: 
\begin{equation}
 \lambda_{\pm}=\pm c_s|\pp|, \quad \lambda_m=\lambda_{T_1,T_2}=\lambda_\parallel=0,
\end{equation}
corresponding to six eigenvectors $\psi_{\bf A}$ where ${{\bf A}}=m,+,-,T_1,T_2,\parallel$. We arrange the eigenvectors to form an orthogonal transformation matrix
\begin{eqnarray}\label{eq:eigenvectors}
\psi^{\cA}_A=\begin{pmatrix} 
1 & 0 & 0 & 0 & 0 & 0\\
0 & {1/ \sqrt{2} } &  -{1/ \sqrt{2} } & 0 & 0 & 0 
\\  
0 & {\php/ \sqrt{2} }  & {  \php/ \sqrt{2} }  & t^{(1)} & t^{(2)} & \bu
   \end{pmatrix},
\end{eqnarray}
where $\php =\pp/|\pp|$ is the unit vector along $\pp$ and $t^{(1)}$ and $t^{(2)}$ are two transverse unit vectors that satisfy
\begin{equation}\label{eq:ti}
t^{(i)}\cdot t^{(j)}=\delta^{ij},\quad t^{(i)}\cdot\hat \pp=0,\quad t^{(i)} \cdot \bu(x)=0.
\end{equation}
Note that the last eigenvector is a consequence of orthogonality
constraint and is not a physical fluctuation mode.  The choice of the
dyad $t^{(i)}(x,q)$ is not unique, and is subject to SO(2) rotations
that are local in both $x$ and $q$ spaces. This local freedom will bring about
additional connections in the confluent derivatives, after we project
$W_{AB}$ onto the slow components.

We go to the basis where  $\LL^{(\p)}$ is diagonal by the orthogonal transformation $M\rightarrow \psi^T M \psi$, upon which $W_{AB}$ transforms to 
\begin{equation}
  \label{eq:W_bAbB}
  \W_{\cA\cB}= \psi_{\cA}^A \W_{AB} \psi_{\cB}^B.
\end{equation}
The spurious components $\W_{\cA \parallel}$, $\W_{\parallel \cB}$,
and $\W_{\parallel \parallel}$ vanish automatically due to the
constraint, Eq.~\eqref{eq:uW=Wu=0}, and we are left with $5\times 5$
matrix $\W_{\cA\cB}$. In the basis, we have
\begin{equation}
[\LL^{(\p)},\W]_{\cA\cB}=(\lambda_{\cA}-\lambda_{\cB})\W_{\cA\cB}, 
\end{equation}
which means that the modes with $\lambda_\cA\neq\lambda_\cB$ are the
fast modes. They average out on the coarse grained time scale
$b_t$ and thus can be neglected. The remaining modes are
  not all independent. In particular,
  \begin{equation}
    \label{eq:W_L}
    W_{++}(x,q) = W_{--}(x,-q) \equiv W_L(x,q)
  \end{equation}
  is the longitudinal mode associated with sound fluctuations.
  The remaining diffusive modes form a $3\times3$ matrix and obey
  $W_{\cA\cB}(x,q)=W_{\cB\cA}(x,-q)$, i.e., only six of these modes are independent.
  These seven independent components, $\W_L$ and $\W_{\cA\cB}$,
  ($\cA,\cB=m, T_1, T_2$), constitute the degrees of freedom in the
new effective kinetic description of fluctuations. Note that the
$3\times 3$ block of $\W_{\cA\cB}\equiv \wtr$ ($\cA,\cB=m, T_1, T_2$)
still contains off-diagonal components, which reflects the fact that
the three modes of $\cA=m, T_1, T_2$ are degenerate and can mix with
each other.

The kinetic equation for the surviving slow components follows
straightforwardly from Eq.~(\ref{eq:kinetic_eq}).  The sound
fluctuation mode completely decouples from other components and
satisfies 
\begin{multline}
(\bu + c_s \php) \cdot\cfd \W_L  =- \gL\pp^2(\W_L-Tw) 
+\left(\left(c_sa_\mu+\partial_{\perp\mu} c_s\right)|\pp|  +(\partial_{\perp\mu} \bu_\nu) \pp^\nu +2 c_s^2 \pp^\lambda \omega_{\lambda\mu}\right)\frac{\partial \W_L}{\partial \pp_\mu}
\\
-\left((1+c_s^2+\dot c_s)\divu +\h_{\mu\nu}\php^\mu \php^\nu+ \frac{1+\left(2-\frac{(\alpha_pTn)^2}{\alpha_m}\right)c_s^2}{c_s} \php\cdot a - \frac{c_s\alpha_pT^2n^2}{\alpha_mw} \hat q\cdot\partial_\perp\alpha \right)\W_L,
\label{eq:Wpm}
\end{multline}
where the sound damping coefficient $\gL$ is given by
\begin{equation}\label{eq:gammaL}
\gL=\gz+\frac{4}{3}\geta+\gp\,,
\end{equation}
and $\gz$, $\geta$ and $\gp$ are defined by Eq.~\eqref{eq:def_gammas}. Here, the confluent derivative of $W_L$ is defined as 
\begin{equation}
  \label{eq:nabla-W-def}
  \cfd_\mu \W_L \equiv \partial_\mu \W_L 
+ \econ^a_{\mu b} q_a\frac{\partial \W_L}{\partial q_b},
\end{equation}
consistent with the fact that $W_L$ behaves as a Lorentz scalar.
Defining
\begin{equation}\label{eq:Npm-Wpm}
  N_L\equiv\frac{W_L}{c_s|q|w},
\end{equation}
such that its equilibrium value, $N_L^{(0)}=T/c_s|q|$, is equal to what one would expect for the distribution function of ``phonons" with the dispersion relation $\omega=c_s |q|$,
Eq.~\eqref{eq:Npm} can be recast into the form that resembles a Boltzmann kinetic equation for phonons,
\begin{eqnarray}\label{eq:Npm}
\mathcal L_L[N_L]&\equiv&\left((\bu + c_s\hat q)\cdot\cfd-\left(\left(c_sa_\mu+\partial_{\perp\mu} c_s\right)|\pp|  +(\partial_{\perp\mu} \bu_\nu) \pp^\nu +2 c_s^2 \pp^\lambda \omega_{\lambda\mu}\right)\frac{\partial}{\partial \pp_\mu}\right) N_L \nn
&=&- \gL\pp^2\left(N_L-\frac{T}{c_s|q|}\right).
\end{eqnarray}
Remarkably, the advection operator $\mathcal L_L[N_L]$ is precisely
equal to the Liouville operator, derived in Ref.~\cite{An:2019rhf}, in
a relativistic kinetic theory of massless particles (phonons), with an
effective inverse metric
$g_{\rm eff}^{\mu\nu}(x)=-c_s^2u^\mu u^\nu+\Delta^{\mu\nu}$ that gives
the local dispersion relation of sound waves $\omega=c_s |q|$. It
should be emphasized that the Liouville operator emerges after
$\partial_\perp \alpha$ terms vanish due to rather nontrivial
cancellations. The simplicity of the collision (relaxation) term in
the right hand side, emerging after cancellation of all the background
gradient terms in Eq.~(\ref{eq:Wpm}), is equally striking.

The diffusive and transverse shear modes, contained in $3\times 3$ matrix $\wtr$, satisfy the matrix equation
\begin{eqnarray}
\bu\cdot \cfd \wtr=-\acomm{\Dtr, \wtr-\wtr^{(0)}}
+(\partial_{\perp\mu} \bu_\nu) \pp^\nu
\nabla_{(q)}^\mu\wtr
-\mcomm{\btr,\wtr} \,,
\label{eq:Nij}
\end{eqnarray}
where
\begin{equation}\label{eq:ij_matrices}
  \begin{aligned}
 &   \Dtr\equiv \left(
      \begin{matrix}
        \gl & 0 \\[2pt]
        0 & \delta_{ij}\gamma_\eta\\
      \end{matrix}
    \right)q^2 , \quad \wtr^{(0)}\equiv Tw\left(
      \begin{matrix}
        \frac{c_pT}{w}  & 0 \\[2pt]
        0 & \delta_{ij}\\
      \end{matrix}
    \right), \quad 
\\
&\btr\equiv \left(
      \begin{matrix}
        \frac{1}{2}(1+2\dot T)\divu & ~\frac{Tn}{\alpha_m}\left(\alpha_p a\cdot t^{(j)}+\frac{1}{w}t^{(j)}\cdot\partial_\perp\alpha\right) \\[2pt]
        \alpha_pTna\cdot t^{(i)} & \frac{1}{2}\divu\,\delta^{ij}+t^{(i)}_\mu t^{(j)}\cdot\partial u^\mu\\
      \end{matrix}
    \right), \quad i=1,2.
  \end{aligned}
\end{equation}
Here we introduced a covariant $\p$-derivative that takes into account the rotation
  of the basis $t^{(i)}(x,q)$ of the transverse modes in $q$ space:
  \begin{equation}
    \label{eq:nabla-q}
    \nabla_{(\p)}^\mu\wtr 
    \equiv \frac{\partial\wtr}{\partial{\pp}_\mu}
    + \left[\widehat\omega^\mu,\wtr\right],\quad
    \mbox{where}\quad  \quad \tcon^{ij}_\mu
    \equiv t_\nu^{(i)}\frac{\partial}{\partial \pp^\mu}t^{(j)\nu}\,,  \quad \tcon^{mm}_\mu=\tcon^{mi}_\mu=\tcon^{im}_\mu=0\,.  
\end{equation}
The confluent derivative in Eq.~(\ref{eq:Nij}) also includes
 additional SO(2) connection $\txcon^{ij}_\mu
    \equiv t_\nu^{(i)}{\partial_\mu}t^{(j)\nu}$, associated with the $x$-dependence of
 the basis vectors $t^{(i)}$: 
  \begin{equation}
    \label{eq:nabla-x-tildeW}
        \cfd_\mu\wtr 
    \equiv \partial_\mu \wtr
  + \econ^a_{\mu b}\, q_a \nabla^b_{(q)} \wtr 
     + \left[\txcon_\mu,\wtr\right].
  \end{equation}
Introducing the rescaled variables
\begin{equation}\label{eq:NW}
  N_{mm}\equiv \frac{W_{mm}}{nT^2}, \quad N_{m \T{i}}\equiv\frac{W_{m \T{i}}}{nT}, \quad N_{\T{i}\T{j}}\equiv\frac{W_{\T{i}\T{j}}}{n},
\end{equation}
and also a Liouville-like operator,
\begin{equation}
\mathcal{L}[\wtr]=\left(\bu\cdot\cfd - (\partial_{\perp\mu} \bu_\nu) \pp^\nu\nabla_{(q)}^\mu\right)\wtr,
\end{equation}
we can simplify Eq. (\ref{eq:Nij}) substantially:
\begin{subequations}\label{eq:LN}
\begin{align}
\mathcal{L}[N_{mm}] =& -2\gl q^2\left(N_{mm}-\frac{c_p}{n}\right)-\frac{n}{w}t^{(i)}\cdot\partial_\perp m(N_{\T{i} m}+N_{m \T{i}}),\label{eq:LN_ss}\\
\mathcal{L}[N_{m \T{i}}] =& -(\geta+\gl) q^2N_{m \T{i}}-\partial^\nu u^\mu t_\mu^{(i)} t_\nu^{(j)} N_{m \T{j}} -\frac{n}{w}t^{(j)}\cdot\partial_\perp m N_{\T{j}\T{i}}+\frac{\alpha_pT^2n}{w}t^{(i)}\cdot\partial_\perp p N_{mm},\label{eq:LN_si}\\
\mathcal{L}[N_{\T{i}\T{j}}] =& -2\geta q^2\left(N_{\T{i}\T{j}}-\frac{Tw}{n}\delta_{ij}\right)-\partial^\nu u^\mu\left(t_\mu^{(i)} t_\nu^{(k)} N_{\T{k}\T{j}}+t_\mu^{(j)} t_\nu^{(k)} N_{\T{i}\T{k}}\right)\nn
&+\frac{\alpha_pT^2n}{w}\partial_\perp^\mu p\left(t^{(i)}_\mu N_{m \T{j}}+ t^{(j)}_\mu N_{\T{i} m}\right)\label{eq:LN_ij},
\end{align}
\end{subequations}
where $\alpha_p=(1-\dot T/c_s^2)/Tn$, given by Eq.~\eqref{eq:2ndordercoeff}. 

The kinetic equations for fluctuations, Eqs. (\ref{eq:Wpm}) and
(\ref{eq:Nij}), are the main results in this section. In the next
section, these equations will be used to isolate short-distance
singularities in the energy-momentum tensor and the charge current,
when we consider two-point correlator contributions to these
observables. This procedure can be done analytically, which allows us
to absorb the short distance singularities into the renormalization of
the equation of state and the first-order transport
coefficients. After the renormalization procedure is carried out
analytically, the resulting renormalized first-order viscous
hydrodynamics with fluctuations  will be well-defined and not suffer from
short-distance ambiguity (i.e., cutoff dependence). These equations
can be then applied to numerical studies of fluctuations in
hydrodynamically evolving systems, such as heavy-ion collisions.


\section{Feedback of fluctuations}
\label{sec:renormalization}
Having studied the dynamics of the fluctuating modes described by the
two-point functions, $W_{AB}(x, \bm q)$, in the previous sections, we
now discuss how these fluctuations affect background hydrodynamic
flow. Hydrodynamics describes the evolution of the average values, or
one-point functions, of hydrodynamic variables, such as $\beps$,
$n$, $u$, or more precisely, by our choice, $m$, $p$, $u$. The equations
governing this evolution are obtained by averaging conservation
equations~(\ref{eq:hydro-stoch}).  However, the evolution of the
one-point functions is affected by the feedback from the higher-point
functions. This is because energy momentum-tensor and charge current
are {\em nonlinear} functions of the fluctuating variables, $\breve m$, $\breve p$ and $\breve u$,
as follows from the constitutive relations,
Eq.~\eqref{eq:constitutive}, as well as the equation state. Therefore
expanding the fluctuating variables in $\phi_A$ inside $\av{T^{\mu\nu}}$
and $\av{J^\mu}$ to quadratic order, we get contributions proportional to 
 \begin{equation}
  \label{eq:G(x)-W}
 \langle \phi_A(x)\phi_B(x)\rangle=G_{AB}(x,y=0)=\int \frac{d^3\p}{(2\pi)^3} W_{AB}(x,\bm q) \equiv G_{AB}(x)\,.
\end{equation}
In this section we discuss two aspects of the fluctuation feedback:
(i) the renormalization of the variables, the equation of state and
the transport coefficients as well as (ii) the time lagged
hydrodynamic response, falling off as a power of time, known as
``long-time tails'', or, equivalently, non-analytic frequency
dependence of the response at low frequencies.

In order to calculate the contribution of the two-point functions, we begin by expanding the energy-momentum tensor and the charge current given in Eq.~\eqref{TandJ}, up to second order in $\phi_A$. Upon averaging over the ensemble,
the linear terms in $\phi_A$ vanish by definition, $\langle \phi_A\rangle=0$, and only the two-point function contributions, expressed in terms of $G_{AB}(x)$, remain.  
Expanding the (bare) equation of state up to second order in fluctuations leads to
\begin{equation}
\begin{aligned}\label{eq:epsn_2nd_expansion}
\eps(\sm,\spp)=&\,\eps(m,p)+\eps_m\dm+\eps_p\dpp+\frac{1}{2}\eps_{mm}
(\dm)^2 +\eps_{mp}\dm\dpp +\frac{1}{2}\eps_{pp}(\dpp)^2+\ldots\,,\\
n(\sm,\spp)=&\,n(m,p)+n_m\dm+n_p\dpp+\frac{1}{2}n_{mm}(\dm)^2 + n_{mp}\dm\dpp+\frac{1}{2}n_{pp}(\dpp)^2+\ldots\,,
\end{aligned}
\end{equation}
where the coefficients of linear terms were already defined in
Eq.~(\ref{eq:dedp-dmdn}). The coefficients of bilinear terms
are third order thermodynamic derivatives and are defined similarly
(see  Appendix~\ref{sec:derivation},
Eqs.~(\ref{eq:coeff_def_3rd})). Similarly to expressions for
second-order thermodynamic derivatives in terms of three independent ones
$c_s$, $c_p$ and $\dot T$ in Eq.~\eqref{eq:2ndordercoeff}, the
third order thermodynamic derivatives can be also expressed in terms
of two independent third order derivatives $\dot c_p$ and $\dot c_s$
as\footnote{Note that there are four independent third order
  derivatives (four independent third order derivatives of entropy), but only two
  are needed in Eqs.~(\ref{eq:3rdordercoeff}). We do not need
  expressions for $\eps_{mp}$ and $n_{mp}$ because they will drop out upon time averaging, as described below.}
\begin{equation}\label{eq:3rdordercoeff}
\begin{gathered}
\eps_{mm}=-\frac{Tn^2}{c_s^2c_p}\left(1-\dot c_p+\dot T-c_s^2+\frac{2c_pT\dot T}{w}\left(1-\frac{\dot T}{c_s^2}\right)\right), \quad \eps_{pp}=-\frac{2\dot c_s}{c_s^4w}, \\
  n_{mm}=-\frac{Tn^3}{c_s^2c_pw}\left(1-\dot c_p+\dot T-\frac{2c_pT\dot T}{c_s^2w}\right), \quad n_{pp}=-\frac{\left(c_s^2+2\dot c_s\right)n}{c_s^4w^2}.
\end{gathered}
\end{equation}
As a result, we obtain the following expansion for $\av{T^{\mu\nu}}$ and $\av{J^\mu}$:
\begin{subequations}
  \begin{eqnarray}\label{eq:avTmunu0}
    \av{\sth T^{\mu\nu}(x)}  
    &=& T^{\mu\nu}(\beps,n,u)  
        +\frac{\eps_{mm}}{2T^2n^2} u^\mu u^\nu G_{mm}(x)+\frac{\eps_{pp}c_s^2}{2} u^\mu u^\nu G_{pp}(x) + \frac{1}{\bw}G^{\mu\nu}(x)\nn
    && +\frac{\eps_m}{wTn}\big(G^{m\mu}(x)\bu^\nu+G^{m\nu}(x)\bu^\mu\big) +\frac{c_s(1+\eps_p)}{w}\big(  G^{p\mu}(x)\bu^\nu+G^{p\nu}(x)\bu^\mu\big),\\
    \av{\sJ^\mu(x)} 
    &=& J^\mu(\beps,n,u)+\frac{n_{mm}u^\mu}{2T^2n^2}G_{mm}(x)+\frac{c_s^2n_{pp}u^\mu}{2}G_{pp}(x) +\frac{n_m}{wTn}G^{m\mu}(x)+\frac{c_sn_p}{w}G^{p\mu}.  \label{eq:avJmu0}
  \end{eqnarray}
\end{subequations}

The mixed term, $G_{mp}(x)\sim\av{\dm\dpp}$, is dropped
because it is a rapidly oscillating component of $G$ whose contribution vanishes
after time averaging, as explained in
Sec. \ref{sec:diagonalization}. Furthermore, we neglect the fluctuations of the
viscous part $\Pi^{\mu\nu}$ relying on the scale
hierarchy $\gamma\p\sim \p/T \ll 1$. The two point
  functions $G_{AB}(x)$ given by solutions of fluctuation kinetic
  equations are nontrivial functionals of the background gradients and
  contain both local and nonlocal terms which
  are associated with renormalization and long-time tails
  respectively.

\subsection{Renormalization of variables and hydrodynamic coefficients}

The locality of the noise in stochastic hydrodynamics is manifested by
the delta functions in Eqs.~\eqref{eq:fdt}.  In the coarse-grained
picture, this singularity is smeared out and the amplitude of the
noise is proportional to $b^{-3/2}$ where $b$ is the size of the fluid
cell. That means taking $b\to0$ requires infinitely large noise.  The
fluid cell must be larger than the microscopic correlation length, say
$T^{-1}$ or $\xi$ whichever larger, for hydrodynamic description to be
valid, but it is otherwise arbitrary. And because it is arbitrary, the
physical results obtained from hydrodynamic equations cannot depend on
the cutoff $b$. 

Because of the infinite (delta function) noise, in our deterministic
formalism, the singularities appear as infinite contributions to
$G_{AB}(x)$, which arise as ultraviolet (UV) divergences in the
integrals over the fluctuation wave-vector $q$ in
Eq.~(\ref{eq:G(x)-W}).
Introducing the UV cutoff $\Lambda=1/b$, we expect that  these $\Lambda$
dependent terms must be absorbed into the renormalized variables, equation of
state and transport coefficients in order for the physics to be cutoff
independent.

This renormalization procedure has been by now well understood in both
nonrelativistic hydrodynamics \cite{Andreev:1978} and relativistic
hydrodynamics without conserved charge \cite{Akamatsu:2018,An:2019rhf}
or in some special cases, such as, e.g., conformal fluids,
\cite{Martinez:2018}.  In this section, we complete this line of
developments by performing the renormalization of hydrodynamics of
arbitrary fluid with
conserved charge in arbitrary backgrounds.  

It must be kept in mind that,
while $\Lambda$ is a high wave-number cutoff from the perspective of
the scale of fluctuations, $q$, it is still small compared to the
microscopic scales, $T$ or $\xi^{-1}$. Therefore even the most
dominant UV divergent contribution to $G_{AB}/w\propto \Lambda^3 T$ is
still a small correction to the average background variables that are
of order $T^4$. However, in practical numerical simulations, these
corrections will introduce a noticeable cutoff dependence, and the
elimination of the cutoff dependence via renormalization is not only a
matter of principle, but also an issue of practical importance.

Our starting point is to identify the physical, or ``renormalized",
fluid velocity $u_R$ and the physical local energy and charge densities $(\eps_R,n_R)$ which are determined by Landau's matching condition
\begin{subequations}
\begin{align}
  \label{eq:<T>u=eu}
  &-\langle \sT^{\mu}_{\nu}\rangle u_R^\nu=\eps_R  u_R^\mu\,,\\
  &-\langle \sJ^\mu\rangle u_{R\mu}=n_R\,.
  \label{eq:<J>u=n}
\end{align}
\end{subequations}
in terms of the ``bare'' variables $u$, $\beps$ and $n$.
Although the fluctuating fluid velocity is properly normalized (i.e. $\su\cdot \su=-1$), the average velocity, $\bu \equiv \av{\su}$, is not since
\begin{eqnarray}
\bu\cdot\bu=-1-\av{\delta u \cdot \delta u}=-1- \frac{1}{\bw^{2}}\N^\mu_\mu(x)\,.
\end{eqnarray}
We define  $u_R$ such that it is normalized, $u_R^2=-1$. Expanding $u_R$ to first order in $G_{AB}$, we obtain \footnote{This expansion is based on the assumption that the two-point function contributions are parametrically smaller than the corresponding bare quantities, due to $\Lambda\ll {\rm min}(T,\xi^{-1})$, as will become clear shortly. Because of this separation of scales, bare quantities that multiply $G_{AB}$ can be simply replaced by their renormalized values.}
\begin{eqnarray}
\bu_{R}^\mu \equiv \frac{\bu^\mu+\frac{\eps_m}{\bw^2Tn} G^{m\mu}(x)+\frac{c_s(1+\eps_p)}{w^2} G^{p\mu}(x)}{\sqrt{1+G^\mu_\mu(x)/\bw^{2}}}\approx \bu^\mu+\frac{\eps_m}{\bw^2Tn} G^{m\mu}(x) +\frac{1+c_s^2}{c_s\bw^2} G^{p\mu}(x) -\frac{\bu^\mu}{2\bw^2 }G^\nu_\nu(x).
\label{eq:umu_redefinition}
\end{eqnarray}
From Eqs.~(\ref{eq:<T>u=eu}) and (\ref{eq:<J>u=n})
we then find
\begin{eqnarray}
\beps_R= \beps + \delta_R\beps,\quad
n_R= n + \delta_R n. \label{eq:endefinition}
\end{eqnarray}
where the fluctuation corrections to local rest frame energy and charge densities are
given by
\begin{subequations}\label{eq:enredefinition}
\begin{align}
\delta_R\beps&= \frac{1}{\bw} \N^\mu_\mu(x)+\frac{\eps_{mm}}{2T^2n^2} \N_{mm}(x)+\frac{c_s^2\eps_{pp}}{2} \N_{pp}(x), 
\\
\delta_R n&= \frac{n}{2w^2}\N^\mu_\mu(x)+\frac{n_{mm}}{2T^2n^2}G_{mm}(x)+\frac{c_s^2n_{pp}}{2}G_{pp}(x).
\end{align}
\end{subequations}
In terms of the $\eps_R$, $n_R$ and $u_R$, we have now the following expressions for $\av{\sT^{\mu\nu}}$ and $\av{\sJ^\mu}$:
\begin{subequations}
  \begin{eqnarray}
    \av{\sT^{\mu\nu}(x)} &=&\beps_R\bu_R^\mu \bu_R^\nu + p(\beps,n)) \bD^{\mu\nu}+\Pi^{\mu\nu}+\frac{1}{\bw} \N^{\mu\nu}(x),\label{eq:avTmunu2}\\
    \av{\sJ^\mu(x)} &=&n_R\bu_R^\mu+\nu^\mu -\frac{n}{w^2}\N^{m\mu}(x)-\frac{c_sn}{w^2}G^{p\mu}(x).\label{eq:avJmu}
  \end{eqnarray}
\end{subequations}

The transformation to physical
variables is not yet complete in Eqs.~(\ref{eq:avTmunu2})
and~(\ref{eq:avJmu}) -- the ``bare'' values $\beps$ and $n$ still
appear in, e.g., $p(\beps,n)$, which will need to be expressed in terms
of physical $\beps_R$ and $n_R$. We shall do this below.

After establishing the expressions for physical energy and charge
densities, our next goal is to determine the physical values of
pressure and transport coefficients. Their physical values differ
from their ``bare values" that appear in the constitutive relations
Eq.~\eqref{eq:constitutive} and Eq.~\eqref{eq:viscous} due to
fluctuations. The fluctuations contain local terms that are zeroth
order (nonvanishing for homogeneous backgrounds) and first order in
gradients. We shall denote these as $\N^{(0)}_{AB}(x)$ and $\N^{(1)}_{AB}(x)$
respectively. The former contributes to the physical value of the
pressure and the latter contributes to the physical values of the
transport coefficients. The remaining parts of $\N_{AB}(x)$, denoted by
$\wt \N(x)_{AB}$, are higher order
in gradients (in fact, as we shall see, they are nonlocal functionals of
hydrodynamic variables):
\begin{equation}
\label{eq:G_decom}
\N_{AB}(x)=\N^{(0)}_{AB}(x)+\N^{(1)}_{AB}(x)+ \wt \N_{AB}(x),
\end{equation}
where the superscripts `$(0)$' and `$(1)$' denote the terms that are
zeroth order and first order in gradient expansion\footnote{Note that
  $\N^{(0)}_{AB}(x)$ still depends on $x$ via terms such as $w(x)$
  however it does not contain any gradient terms such as $\del_\mu u$
  or $\del_\mu \alpha$ and it does not vanish in a homogeneous
  background. $\N^{(1)}_{AB}(x)$  terms are explicitly linear in
  gradients and do vanish in a homogeneous background. }. Similarly
since $\delta_R\beps$ and $\delta_R n$ in
Eq.~(\ref{eq:enredefinition}) are linear combinations of
$G_{AB}(x)$, these quantities can be also expanded:
\begin{equation}
  \label{eq:delta^01wt}
  \delta_R\beps
=\delta_R^{(0)}\beps+\delta_R^{(1)}\beps+ \wt  \delta_R\beps,\quad
  \delta_R n
=\delta_R^{(0)} n+\delta_R^{(1)}n+ \wt  \delta_Rn,
\end{equation}
where expressions for $\delta_R^{(0)}(\beps,n)$, $\delta_R^{(1)}(\beps,n)$ and
$\wt\delta_R(\beps,n)$ are the same as  $\delta_R(\beps,n)$ in
Eq.~(\ref{eq:enredefinition}) with $\N_{AB}$ replaced with
$\N^{(0)}_{AB}$, $\N^{(1)}_{AB}$ and $\wt G_{AB}$ respectively.

By substituting this gradient expansion, Eqs. \eqref{eq:G_decom} and~(\ref{eq:delta^01wt})  into
Eqs. \eqref{eq:avTmunu2} and \eqref{eq:avJmu} we can identify the
physical values of pressure and transport coefficients by collecting
terms zeroth order in gradients into physical (renormalized) pressure
$p_R$ and terms first order in gradients into physical (renormalized) values of kinetic coefficients:
\begin{subequations}\label{eq:constitutive-R-TJ}
\begin{align}
\av{\sT^{\mu\nu}(x)} 
&=\beps_R\bu_R^\mu \bu_R^\nu + p_R
    \bD^{\mu\nu}+\Pi_R^{\mu\nu}+  \wt T^{\mu\nu} ,\label{eq:avTmunu-R}\\
\av{\sJ^\mu(x)} 
&=n_R\bu_R^\mu+\nu_R^\mu 
 + \wt J^{\mu}.\label{eq:avJmu-R}
\end{align}
\end{subequations}
where the zeroth-order terms in gradient expansion are given by 
\begin{multline}
  \label{eq:p_R}
  p_R(\beps_R,n_R)\Delta^{\mu\nu} = 
p(\beps,n) \Delta^{\mu\nu} +  \frac{1}{\bw} {\N^{\mu\nu}}^{(0)}(x)
\\= \left( p(\beps_R,n_R) 
- \left(\frac{\partial p}{\partial\beps}\right)_n \delta_R^{(0)}\beps
- \left(\frac{\partial p}{\partial n}\right)_\beps \delta_R^{(0)}n
\right)\Delta^{\mu\nu}
+ \frac{1}{\bw} {\N^{\mu\nu}}^{(0)}(x)\,,
\end{multline}
and the first-order terms are given by
\begin{subequations}\label{eq:Pi_nu_R}
\begin{align}
  \label{eq:Pi_R}
  \Pi_R^{\mu\nu} &= \Pi^{\mu\nu} - \left(
\left(\frac{\partial p}{\partial\beps}\right)_n \delta_R^{(1)}\beps
+ \left(\frac{\partial p}{\partial n}\right)_\beps \delta_R^{(1)}n
\right)\Delta^{\mu\nu}
+ \frac{1}{\bw} {\N^{\mu\nu}}^{(1)}(x)\,,\\
   \label{eq:nu_R}
   \nu_R^\mu &= \nu^\mu 
 -\frac{n}{w^2}{\N^{m\mu}}^{(1)}(x)
 -\frac{c_sn}{w^2}{\N^{p\mu}}^{(1)}(x).
\end{align}
\end{subequations}
The remaining, i.e.,  higher-order in gradients (and nonlocal), contributions to constitutive equations are given by
\begin{subequations}\label{eq:wt-TJ}
\begin{align}
  \label{eq:wt-Tmunu}
  \wt T^{\mu\nu} &=  -\left(
\left(\frac{\partial p}{\partial\beps}\right)_n \wt\delta_R\beps
+ \left(\frac{\partial p}{\partial n}\right)_\beps \wt\delta_Rn
\right)\Delta^{\mu\nu}
+ \frac{1}{\bw} {\wt\N^{\mu\nu}}(x)\nn
&=\frac{1}{2w}\left((1-\dot c_p)\frac{w}{c_pT}\wt G_{mm}(x)+\left(c_s^2-\dot T+2\dot c_s\right)\wt G_{pp}(x)-\left(c_s^2+\dot T\right){\wt G}^\lambda_\lambda(x)\right)\bD^{\mu\nu}+\frac{1}{w} {\wt G^{\mu\nu}}(x)\,,
\\
  \label{eq:wt-Jmu}
  \wt J^\mu &= -\frac{n}{w^2}\wt\N^{m\mu}(x)-\frac{c_sn}{w^2}\wt\N^{p\mu}(x).
\end{align}
\end{subequations}

Let us now work out the explicit expressions for the physical pressure and transport coefficients. According to Eq.~\eqref{eq:G(x)-W}, the corresponding decomposition of Eq.~\eqref{eq:G_decom} in phase space is given by
 \begin{equation}\label{eq:W_decom}
W_{AB}(x,q)=W_{AB}^{(0)}(x,q)+W^{(1)}_{AB}(x,q)+ \wt W_{AB}(x,q).
\end{equation}
The zeroth-order contribution $\N_{AB}^{(0)}$
follows from the equilibrium solution to the fluctuation
evolution equations given by Eq.~(\ref{eqvalue}): 
\begin{equation}
\W^{(0)}_{AB}(x, \p)=\text{diag}\left(c_pT^2,Tw,Tw\bD_{\mu\nu}\right)\,.\label{eq:W0}
\end{equation}
Since $\W^{(0)}$ does not depend on $\p$, the integration over $\p$ is
divergent. We regularize this integral by the wave-number cutoff $\p<\Lambda$. 
\begin{eqnarray}
\N^{(0)}_{AB}(x) = \int \frac{d^3q}{(2\pi)^3}\W^{(0)}(x,q)
=\frac{\Lambda^3}{6\pi^2}\text{diag}\left(c_pT^2,Tw,Tw\bD_{\mu\nu}\right).
\label{eq:Lambda3}
\end{eqnarray}
Though the tensor part of the two-point function, $\N^{(0)\mu\nu}(x)$
appearing in Eq.~(\ref{eq:p_R}) is cutoff-dependent, it is
proportional to $\Delta^{\mu\nu}$, and thus is absorbed into the
definition of the physical (renormalized) pressure.
Combining this contribution with the contributions from the terms containing
$\delta_R^{(0)}\beps$ and $\delta_R^{(0)}n$ in
Eq.~(\ref{eq:p_R}) we find for the renormalized, i.e., physical, pressure:
\begin{eqnarray}\label{eq:p-renormalized-Lambda}
  p_R(\eps_R, n_R)&=&p(\eps_R,n_R)+\frac{1-3\left(1-\frac{\eps_m}{2Tn}\right)c_s^2}{3w}{G^{(0)}}^\mu_\mu\nn
  &&+\frac{c_s^2w}{2T^3n^4}\left(n_m\eps_{mm}-\eps_m n_{mm}\right)G^{(0)}_{mm}+\frac{c_s^4w}{2Tn^2}\left(n_m\eps_{pp}-\eps_mn_{pp}\right)G^{(0)}_{pp}\nn
  &=& p(\eps_R,n_R)+\frac{2-3\left(c_s^2+\dot T \right)}{6w}{G^{(0)}}^\mu_\mu +\frac{1-\dot c_p}{2c_pT}G^{(0)}_{mm}+
 \frac{c_s^2-\dot T+2\dot c_s}{2w}G^{(0)}_{pp}
\nn
&=&p(\eps_R,n_R)+\frac{T\Lambda^3}{6\pi^2}\left((1-c_s^2-2\dot T+\dot c_s)+\frac{1}{2}(1-\dot c_p)\right). 
\end{eqnarray}
where we used 
\begin{equation}\label{eq:1-cpdot}
  n_m\eps_{mm}-\eps_m n_{mm}=\frac{T^2n^4}{c_s^2c_pw}(1-\dot c_p), \quad n_m\eps_{pp}-\eps_mn_{pp}=\frac{Tn^2}{c_s^4w^2}(c_s^2-\dot T+2\dot c_s),
\end{equation}
which can be derived by using Eq.~\eqref{eq:2ndordercoeff} and
\eqref{eq:3rdordercoeff}. This procedure of defining the physical
pressure that combines ``bare'' pressure with the effects of equilibrium fluctuations is
similar to the standard renormalization procedure in quantum field
theory.  Having performed the renormalization of
  hydrodynamic variables and the equation of state, in what
follows, for notational simplicity, we will drop the subscript $R$
on hydrodynamic variables $\epsilon_R$, $n_R$ and $u_R$ and
thermodynamic functions such as $p_R$.

We now turn to the first-order terms in the gradient expansion
given by Eq.~(\ref{eq:Pi_nu_R}). Since these terms
are linear in gradients, they must be combined with the ``bare''
transport terms into the physical
transport terms. It may seem that this procedure, similar to
renormalization of pressure, is guaranteed to succeed. It indeed does,
but this is not trivial because not all gradient (transport) terms are
allowed by second law of thermodynamics. The fact that only those that
are allowed
arise from fluctuation emerges after delicate cancellations and is a
nontrivial test of the conceptual validity of the framework we develop.

Let us begin with calculating $\W^{(1)}_{AB}(x,\p)$. This calculation essentially follows the same steps given in Ref.~\cite{An:2019rhf} but in this case there is an additional mode associated with conserved charge. We begin with inserting the decomposition Eq.~\eqref{eq:W_decom} into our main kinetic equation given in Eqs. \eqref{eq:Npm} and \eqref{eq:Nij}. This substitution leads to an equation for $W_{\cA\cB}^{(\rm neq)}(x,q)$ which we then expand to first order in gradients of the background flow.
Because the kinetic equation already contains gradients of the leading term $\W^{(0)}_{\cA\cB}(x)$, we can use the ideal equations of motion to convert the time derivatives into spatial derivatives:
\begin{equation}
\begin{aligned}
\bu\cdot\partial(T\bw)&=-\left(1+c_s^2+\dot T\right)Tw\divu, \quad \bu\cdot\partial(c_pT^2)=-c_pT^2(\dot c_p+2\dot T)\divu\,\\
\partial_{\perp\mu}(T\bw)&=-Tw\left(\frac{1+2c_s^2}{c_s^2}+\left(1-\frac{\dot T}{c_s^2}\right)^2\frac{c_pT}{w}\right)a_\mu-T^2n\left(1+\left(1-\frac{\dot T}{c_s^2}\right)\frac{c_pT}{w}\right)\partial_\perp\alpha\,.
\end{aligned}
\end{equation}
In deriving these, we use thermodynamic relations given in Appendix \ref{sec:derivation}. Keeping only the terms that are linear in background gradients, the equations for $\W_{\cA\cB}^{(1)}$ can be solved as
\begin{equation}
\begin{gathered}
\W^{(1)}_L(x,\pp)= \frac{T\bw}{\gL\pp^2}\left((\dot T-\dot c_s)\divu -\h_{\mu\nu}\hat \pp^\mu \hat \pp^\nu+ \frac{c_sTn}{w}\hat q\cdot\partial_\perp\alpha\right),
\\
\W^{(1)}_{mm}(x,\pp)= \frac{c_pT^2}{2\gl q^2}\left(\dot c_p-1\right)\divu\,, \quad   W^{(1)}_{\T{i}m}(x,\pp) = W^{(1)}_{m\T{i}}(x,\pp)= \frac{c_pT^3n/w}{(\geta+\gl) q^2}t^{(i)}\cdot\partial_\perp\alpha,
\\
\W^{(1)}_{\T{i}\T{j}}(x,\pp) = \frac{T\bw}{2\geta\pp^2}\left(\left(c_s^2+\dot T\right)\divu\,\delta^{ij}-2\h^{\mu\nu}t^{(i)}_\mu t^{(j)}_\nu\right).
   \label{eq:N1pmij}
\end{gathered}
\end{equation}
Note that these expressions are given in the  $(\cA,\cB)$ basis where $\LL^{(\p)}$ is diagonal and we need to convert them back into the $(A,B)$ basis:
\begin{equation}\label{eq:WAB-WcAcB}
\begin{aligned}
\W_{AB}&=\psi^{\cA}_A \W_{\cA\cB} \psi^{\cB}_B
=\begin{pmatrix} 
W_{mm} & W_{m p} & W_{m\nu}\\ 
W_{pm} & W_{pp} & W_{p\nu} \\
W_{\mu m} & W_{\mu p}  & W_{\mu\nu}
\end{pmatrix} \\
&=\begin{pmatrix} 
W_{mm} & 0 & W_{m\T{j}}t_\nu^{(j)}\\ 
0 & \frac{1}{2}(\W_{++} + \W_{--}) & \frac{1}{2}(\W_{++} - \W_{--}) \hat \pp_\nu \\
W_{\T{i}m}t_\mu^{(i)} & \frac{1}{2}(\W_{++} - \W_{--})\php_{\mu}  & ~\frac{1}{2}(\W_{++} + \W_{--}) \php_{\mu} \php_{\nu} +\W_{\T{i}\T{j}}t_\mu^{(i)}t_\nu^{(j)}
\end{pmatrix},
\end{aligned}
\end{equation}
which finally gives $W^{(1)}_{AB}(x,q)$ in components,
\begin{equation}
\begin{aligned}\label{eq:W1}
&\W^{(1)}_{mm}(x,\pp)=\frac{c_pT^2}{2\gl q^2}\left(\dot c_p-1\right)\divu, \quad \W^{(1)}_{pp}(x,\pp)=\frac{T\bw}{\gL \pp^2}\left((\dot T-\dot c_s)\divu -\h_{\mu\nu}\php^\mu \php^\nu \right), \\
&\W^{(1)}_{m\mu}(x,\pp)=\W^{(1)}_{\mu m}(x,\pp)= \frac{c_pT^3n/w}{(\geta+\gl) q^2}t_\mu^{(i)}t^{(i)}\cdot\partial_\perp\alpha, \quad \W^{(1)}_{p\mu}(x,\pp)=\W^{(1)}_{\mu p}(x,\pp)=\frac{c_sT^2n}{\gL\pp^2}\hat q^\mu\hat q\cdot\partial_\perp\alpha,\\
&\W^{(1)}_{\mu\nu}(x,\pp)=\frac{T\bw}{\gL \pp^2}\left((\dot T-\dot
  c_s)\divu -\h_{\lambda\kappa}\php^\lambda\php^\kappa \right)\php_{\mu} \php_{\nu} +\frac{T\bw}{2\geta \pp^2}\left(\left(c_s^2+\dot T\right)\divu\,\dtr_{\mu\nu}- 2\h^{\lambda\kappa}\dtr_{\lambda\mu}\dtr_{\kappa\nu}\right),
\end{aligned}
\end{equation}
where $\dtr_{\mu\nu} = \sum_{i=1}^2t^{(i)}_\mu t^{(i)}_\nu=\bD_{\mu\nu}-\php_{\mu} \php_{\nu}$. 
Then, the corresponding $G^{(1)}_{AB}(x)$ after $q$ integration are given by
\begin{subequations}\label{eq:G(1)}
  \begin{align}
&\N^{(1)}_{mm}(x)=\frac{c_pT^2\Lambda}{4\pi^2\gl}\left(\dot c_p-1\right)\divu, \quad \N^{(1)}_{pp}(x)=-\frac{T\bw\Lambda}{6\pi^2\gL}\left(1-3\dot T+3\dot c_s\right)\divu \label{eq:G1-1}
,\\
&\N^{(1)}_{m\mu}(x)=
\frac{(c_pT^3n/w)\Lambda}{3\pi^2(\geta+\gl)}\partial_{\perp\mu}\alpha,
\quad
\N^{(1)}_{p\mu}(x)=\frac{c_sT^2n\Lambda}{6\pi^2\gL}\partial_{\perp\mu}\alpha, \label{eq:G1-2}\\
&\N^{(1)}_{\mu\nu}(x)=-\frac{T\bw\Lambda}{6\pi^2\gL}\left(\left(\frac{1}{5}-\dot T+\dot c_s\right)\divu\bD_{\mu\nu} +\frac{2}{5} \h_{\mu\nu}\right)-\frac{T\bw\Lambda}{60\pi^2 \geta }\left((2-10(c_s^2+\dot T))\divu\bD_{\mu\nu} +14\h_{\mu\nu}\right). \label{eq:G1-3}
\end{align}
\end{subequations}
Finally we substitute the above expressions for $G^{(1)}$ into
Eq.~(\ref{eq:Pi_nu_R}). The resulting contributions are linear in the
gradients and have the same form as ``bare'' viscous terms in $\Pi^{\mu\nu}$
and diffusion term in $\nu^\mu$. Therefore they can be absorbed into the definitions of
viscosities $\eta$, $\zeta$ and conductivity $\lambda$. After
straightforward computation, we obtain the renormalized transport
coefficients as
\begin{eqnarray}\label{eq:eta_R}
\eta_R&=&\eta+\frac{T\Lambda}{30\pi^2}\left(\frac{1}{\gL}+\frac{7}{2\geta} \right)  ,\\
\label{eq:zeta_R}
\zeta_R&=& \zeta+\frac{T\Lambda}{18\pi^2}\left(\frac{1}{\gL}(1-3\dot T+3\dot c_s)^2+\frac{2}{\geta}\left(1-\frac{3}{2}(\dot T+c_s^2) \right)^2+\frac{9}{4\gl}(1-\dot c_p)^2 \right)   ,\\\label{eq:lambda_R}
\lambda_R&=&\lambda+\frac{T^2n^2\Lambda}{3\pi^2w^2}\left(\frac{c_pT}{(\geta+\gl)w}+\frac{c_s^2}{2\gL}\right).
\end{eqnarray}
A couple of comments are in order. First, all the gradients appearing
in the expansion of  $G^{(1)}$ are matched by the gradients appearing
in the first-order terms in the constitutive equations,
$\Pi^{\mu\nu}$ and $\nu^\mu$. For $\Pi^{\mu\nu}$, this is a simple
consequence of the fact that, by construction, $\Pi^{\mu\nu}$ involves
all gradients allowed by Lorentz symmetry, so nothing else could have appeared
in Eqs.~(\ref{eq:G1-1}) or~(\ref{eq:G1-3}). However, this is less
trivial in the case of the corrections to $\nu^\mu$. This is because there are {\em two} linearly independent
gradient terms allowed by Lorentz symmetry alone, e.g., $\del_\mu \alpha$ and $\del_\mu p$,
and, naively, any their linear combination could have appeared in the
expression for $G^{(1)}$ in Eqs. \eqref{eq:G1-2}.  However, precisely
$\del_\mu \alpha$ appears in Eqs.~(\ref{eq:G1-2}), which allows us to
absorb the fluctuation contribution into $\lambda_R$. Any other linear
combination would require additional kinetic coefficient to absorb
it. However, the second law of thermodynamics only allows the gradient
$\del_\mu \alpha$ to appear in $\nu^\mu$ in order to
guarantee the semi-positivity of entropy production rate. The way this
constraint is respected by fluctuation contributions appears to be
highly nontrivial, relying on delicate cancellations that  result in
rather elegant thermodynamic identities given by
Eq.~\eqref{eq:1-cpdot}. Of course, we can view this as one of the many nontrivial
checks of the consistency of this approach and the validity of the
calculations.

Second, in a similarly remarkable deference to the second law of
thermodynamics manifested in delicate cancellations, the correction to the bulk viscosity given in
Eq.~\eqref{eq:zeta_R} is nonnegative. Also, as expected, but similarly
achieved through nontrivial cancellations, the
fluctuation corrections vanish
in the conformal limit, where $c_s^2=1/3$ and $\eps=3p$, when bulk
viscosity must vanish.

\subsection{Long time tails}
\label{sec:long-time-tails}
After all constitutive equations are expressed in terms of the
physical, i.e., renormalized, variables, pressure and transport
coefficients, the remaining contributions, denoted by $\wt T^{\mu\nu}$
are cutoff independent. This is very similar to renormalization in
quantum field theory, and it works for a similar reason -- the locality of the
first-order hydrodynamics (similar to the locality of quantum field theory
Lagrangian). On a more technical level, the gradient expansion in
$W_{AB}$ is accompanied by the expansion in $1/q^2$. This can be
traced back to the power-counting scheme in which $k\sim q^2$. The
terms of order $k^2$ are accompanied by $1/q^4$ leading to
convergent integrals in $\wt\N_{AB}$.

Thus, expressed in terms of physical quantities, the constitutive
equations~(\ref{eq:constitutive-R-TJ}) do not contain
UV divergences which could lead to cutoff dependence. 
Together with 
the conservation equations
\begin{subequations}\label{eq:conservation-TJ}
\begin{align}
  &\partial_\mu\av{\sT^{\mu\nu}(x)} =0\,, \label{eq:d<TR>=0}\\
  &\partial_\mu\av{\sJ^{\mu}(x)} =0\,,  \label{eq:d<JR>=0}
\end{align}
\end{subequations}
and the fluctuation evolution equations~\eqref{eq:Npm}
and~\eqref{eq:Nij}, they now form a closed set of cutoff-independent,
deterministic equations that describe the evolution of the background
flow, including the feedback of the fluctuating modes $\wt W$. 

In principle this coupled system of equations can be solved
numerically and nonlocal effects of long time tails in an arbitrary
background can be studied. We leave such a numerical study for future
work. Instead, for the remainder of this section we will describe
important analytical properties of the long-time tails in simple backgrounds by solving the fluctuation evolution equations \eqref{eq:Npm}. 

A quick look at the evolution equations, \eqref{eq:Nij} and \eqref{eq:Npm} leads to the following ``impressionistic" expression for the nonequilibrium part of the Wigner function:
\begin{equation}
  \label{eq:Wneq}
   W^{(\rm neq)}\equiv W-W^{(0)} \sim 
\frac{\partial f}{\gamma\pp^2+i( u + v)\cdot  k+\del f}\,
\end{equation}
where $v=\pm c_s\hat\pp$ or $0$ depending on which mode we are
considering and $\gamma$ and $\del f$ are schematic notations for the relaxation rate coefficients and terms linear in background gradients respectively.  Note that $k \sim \del$ and $u\cdot k=\omega$ is the frequency.
After subtracting the term linear in the background gradients, which is absorbed into the definitions of renormalized transport coefficients, we obtain a schematic expression for the finite part of the Wigner function:
\begin{equation}
  \label{eq:tildeW}
  \wt \W
  \sim \frac{\partial f}{\gamma\pp^2+i( u + v)\cdot  k+\del f}-\frac{\partial f}{\gamma\pp^2}
  \sim
\frac{( u + v)\cdot k}{\gamma\pp^2+i( u + v)\cdot  k+\del f}\,
\frac{\del f}{\gamma\pp^2}
\end{equation}
This procedure could be viewed a a subtraction scheme that regulates
the phase space integral of the fluctuation modes where the local (and
instantaneous) short distance term is subtracted. The integration over
$\pp$ leads to $\wt\N \sim k^{1/2}\del f/\gamma^{3/2}\sim k^{3/2}$
which is a nonlocal functional of the gradients
\cite{Andreev:1978}. Notice that $k^{3/2}$ in terms of gradient
expansion lies in between $k$ (first order, viscous terms) and $k^2$
(second order terms) After Fourier transformation these terms lead to
power-law corrections which correspond to the long-time tails.  

To be more quantitative, let us consider a special case and focus on
the non-analytic $\omega$ dependence, by taking spatial $k$ to zero
for simplicity. This means that we only keep the $k$ dependence for
the background gradient term $\del f$ that is in the numerator of
Eq.~\eqref{eq:Wneq} which is consistent with the order of gradient
expansion that we are working with.  In other words we are looking at
the frequency dependence of the transport coefficients. From
Eq.~\eqref{eq:tildeW} we see that  the frequency dependence can be
expressed as
\begin{equation}
  \label{eq:Womega-q2}
 \wt W(x,q)=W^{(1)}(x, q \,)\Big|_{\gamma q^2\to \gamma q^2-i\omega}-W^{(1)}(x,q). 
\end{equation}  
The contribution of  the two-point functions to the constitutive relation for the charge current is given in 
Eq.~\eqref{eq:wt-Jmu}. We can calculate the relevant $\wt W(x,q)$ by using the substitution, Eq.~\eqref{eq:Womega-q2}, in Eq.~\eqref{eq:W1}. By plugging the resulting expression into Eq.~\eqref{eq:wt-Jmu}, we obtain 
\begin{equation}
\begin{aligned}
\lambda(\omega)\partial_{\perp}^\mu\alpha\equiv&\,\lambda\partial_{\perp}^\mu\alpha +\frac{n}{w^2}\wt G^{m\mu}(x)+\frac{c_sn}{w^2}\wt G^{p\mu}(x)\\
=&\,\lambda\partial_{\perp}^\mu\alpha +i\omega\frac{c_pT^3n^2}{w^3}\partial_{\perp}^\nu\alpha\int_q\frac{\Delta_{\mu\nu}-\hat q_\mu\hat q_\nu}{((\geta+\gl)q^2-i\omega)(\geta+\gl)q^2}+i\omega\frac{c_s^2T^2n^2}{w^2}\partial_{\perp}^\nu\alpha\int_q\frac{\hat q_\mu\hat q_\nu}{(\gL q^2-i\omega)\gL q^2},\\
\end{aligned}
\end{equation}
from which we find the frequency dependent conductivity, $\lambda(\omega)$, to be
\begin{equation}\label{eq:lambda_omega}
\begin{aligned}
\lambda(\omega)=&\,\lambda -\omega^{1/2}\frac{T^2n^2}{w^2}\frac{(1-i)}{6\sqrt{2}\pi}\left(\frac{c_pT}{(\geta+\gl)^{3/2}w} +\frac{c_s^2}{2\gL^{3/2}}\right)\,.
\end{aligned}
\end{equation}
Here, $\lambda$ denotes the \textit{renormalized} value of the zero
frequency conductivity. This result is consistent with the already known result for the special case of a conformal, boost invariant plasma with conserved charge given in Eq.~(50b) in Ref. \cite{Martinez:2018}.

The frequency dependent viscosities can be computed in the same way. The fluctuation contributions to the viscous tensor is:
\begin{equation}
\begin{aligned}
\label{eq:piomega}
\Pi^{\mu\nu}(\omega)&\equiv-2\eta(\omega)\left(\h^{\mu\nu}-\frac{1}{3} \Delta^{\mu\nu}\divu\right)-\zeta(\omega)\divu\Delta^{\mu\nu}\\
&\equiv\,\Pi^{\mu\nu} +\frac{1}{\bw} \wt\N^{\mu\nu}(x)+\frac{1}{2w}\left(\frac{(1-\dot c_p)w}{c_pT}\wt G_{mm}(x)+\left(c_s^2-\dot T+2\dot c_s\right)\wt G_{pp}(x)-\left(c_s^2+\dot T\right){\wt G}^\lambda_\lambda(x)\right)\bD^{\mu\nu},\\
\end{aligned}
\end{equation}
where $\Pi^{\mu\nu}$ stands for $\Pi^{\mu\nu}(\omega=0)$. After substituting the $\omega$ dependence in Eq.~\eqref{eq:Womega-q2} in Eq.~\eqref{eq:piomega} we obtain
\begin{equation}
\begin{aligned}
\Pi^{\mu\nu}(\omega)=&\,\Pi^{\mu\nu} +i\omega T\int_q\left\{\frac{\left((\dot T-\dot c_s)\divu -\h_{\lambda\kappa}\php^\lambda\php^\kappa \right)\php^{\mu} \php^{\nu}}{(\gL q^2-i\omega)\gL q^2} 
+\frac{\left(c_s^2+\dot T\right)
\divu\,\dtr^{\mu\nu}- 2\h^{\lambda\kappa}\dtr_{\lambda}^\mu\dtr_{\kappa}^\nu}{(2\geta q^2-i\omega)2\geta q^2}\right\}
\\
&+\frac{i\omega T}{2}\bD^{\mu\nu}\int_q\left\{-\frac{(1-\dot c_p)^2\theta}{(2\gl q^2-i\omega)2\gl q^2}+\frac{\left(c_s^2-\dot T+2\dot c_s\right)\left((\dot T-\dot c_s)\theta-\theta_{\mu\nu}\hat q^\mu\hat q^\nu\right)}{(\gL q^2-i\omega)\gL q^2}\right.
\\
&\left.~~~~~~~~~~~~~~~~~~~~-\left(c_s^2+\dot T\right)\left[\frac{(\dot T-\dot c_s)\theta-\theta_{\lambda\kappa}\hat q^\lambda\hat q^\kappa}{(\gL q^2-i\omega)\gL q^2}+\frac{2\left(c_s^2+\dot T\right)\theta-2\theta_{\lambda\kappa}\dtr_{\lambda\kappa}}{(2\geta q^2-i\omega)2\geta q^2}\right]\right\}
\end{aligned}
\end{equation}
from which find the frequency dependent viscosities,
\begin{equation}\label{eq:viscosities_omega}
\begin{aligned}
\eta(\omega)=&\,\eta-\omega^{1/2}T\frac{(1-i)}{60\sqrt{2}\pi}\left(\frac{1}{\gL^{3/2}}+\frac{7}{(2\geta)^{3/2}}\right),\\
\zeta(\omega)=&\,\zeta-\omega^{1/2}T\frac{(1-i)}{36\sqrt{2}\pi}\left(\frac{1}{\gL^{3/2}}\left(1-3\dot T+3\dot c_s\right)^2+\frac{4}{(2\geta)^{3/2}}\left(1-\frac{3}{2}(\dot T+c_s^2) \right)^2+\frac{9}{2(2\gl)^{3/2}}\left(1-\dot c_p\right)^2\right).
\end{aligned}
\end{equation}
Here, $\eta$ and $\zeta$ denote the \textit{renormalized} values of the zero frequency viscosities.


\section{Fluctuations near the critical point}
\label{sec:fluctnearcp}
\providecommand\phineq{\varphi^{\rm (neq)}}
\providecommand\bmphieq{\bm\varphi^{(0)}}
\providecommand\phieq{\varphi^{(0)}}

In the preceding sections we saw that kinetic coefficients,
$\zeta$, $\eta$ and $\lambda$ receive contributions from
fluctuations. These contributions are
dominated by the fluctuations at the cutoff scale $\Lambda$ and
therefore depend on the cutoff.

In this section we consider the physics of fluctuations at the
critical point. The main feature of the critical point is that the
equilibrium correlation length of the fluctuations, $\xi$, becomes
infinite. To maintain the separation between the hydrodynamic scales
$L\sim k^{-1}$ and microscopic scales, such as $\xi$, we must limit
the domain of applicability of hydrodynamic description to wave-vectors
$k\ll\xi^{-1}$. However, as emphasized in
Ref.~\cite{Stephanov:2018hydro+}, this does not mean that
hydrodynamics applies until $k\sim\xi^{-1}$.  Instead, hydrodynamics
breaks down before $k$ reaches that limitation. Hydrodynamics breaks
down when the frequency of the fastest hydrodynamic mode (the sound,
with $\omega\sim c_s k$) reaches the rate of the relaxation of the
slowest non-hydrodynamic mode.  Near the critical point this rate
vanishes much faster than $\xi^{-1}$.

The slowest non-hydrodynamic variable at the critical point is the
fluctuation of the slowest hydrodynamic mode (diffusive mode $m$),
given by $N_{mm}$. The relaxation rate depends on $q$ and equals
$2\gl q^2$ for $q\ll\xi^{-1}$. Because the contribution of the
fluctuations to pressure and kinetic coefficient is UV divergent, it
is dominated by the modes near the cutoff, which in the case of the
critical point is effectively $\Lambda\sim\xi^{-1}$. Thus the
characteristic rate of non-hydrodynamic relaxation, $\Gamma_\xi$, is
of order $\gl \xi^{-2}$. Together with the fact that $\gl$ vanishes as
a power of $\xi$, i.e., to a good approximation $\gl\sim \xi^{-1}$,
\footnote{This can be easily estimated from
  Eq.~(\ref{eq:lambda_R}). The contribution of fluctuations which
  dominates at the critical point is in the term proportional to
  $c_p$, i.e., $\lambda_R\sim \Lambda c_p$. Given that $c_p\sim \xi^2$
  and $\Lambda\sim \xi^{-1}$, we find $\lambda\sim \xi^1$ and
  $\gamma_\lambda \sim \lambda/c_p\sim \xi^{-1}$. We neglected the
  critical exponent $\eta_x$ ($c_p\sim\xi^{2-\eta_x}$) and the
  divergence of the shear viscosity $\eta\sim \xi^{x_\eta}$ (an error
  of less than 10\%). Taking those into account, we would obtain the
  exact relation for the exponent $x_\lambda$, defined by
  $\lambda\sim\xi^{x_\lambda}$: $x_\lambda=d-2-x_\eta-\eta_x$
  (cf. Ref.~\cite{Hohenberg:1977}). Since
  $\Gamma_\xi\sim\gamma_\lambda\xi^{-2}$, the standard dynamical
  critical exponent $z$ defined as $\Gamma_\xi\sim\xi^{-z}$ is related
  to $x_\lambda$ as $z=4-\eta_x-x_\lambda$.\label{fn:z}} 
we find
that the hydrodynamic breaks down already when the frequency reaches
$\omega\sim \xi^{-3}$. For the sound modes this corresponds to
$k\sim\xi^{-3}$, much earlier than $\xi^{-1}$.

To extend hydrodynamics past $k\sim \xi^{-3}$ we need to include the
slowest non-hydrodynamic mode, which is the idea behind Hydro+
\cite{Stephanov:2018hydro+}. In our notations this mode (or modes,
labeled by index $q$) is $N_{mm}$. In this section we intend to show
that in the regime $k>\xi^{-3}$ our formalism reproduces Hydro+. This
is a nontrivial check because Hydro+ formalism was derived in
Ref.~\cite{Stephanov:2018hydro+} using a completely different approach
by considering a generalized entropy which depends on the
non-hydrodynamic variables (2PI entropy).

The formalism of Hydro+, while
extending ordinary hydrodynamics beyond the scales $k\sim\xi^{-3}$, in
turn, also breaks down well before $k$ reaches $k\sim\xi^{-1}$. The breakdown
occurs when the frequency reaches the relaxation rate $\Gamma_\xi'$ of the
next-to-slowest nonhydrodynamic mode. This mode (or modes) are the fluctuations of
velocity transverse to the wave-vector. This relaxation rate is of
order $\geta q^2$ at $q\ll \xi^{-1}$. Again, the dominant contribution
comes from modes at $q\sim\xi^{-1}$ and, since $\geta$ to a good
approximation can be treated as finite at the critical point
\cite{Hohenberg:1977}, Hydro+ breaks down when frequency reaches
$\omega\sim\Gamma_\xi'\sim \xi^{-2}$, which for the sound modes corresponds to
$k\sim \xi^{-2}$. Near the critical point this scale is still much
lower than $\xi^{-1}$. 

In our formalism the next-to-slowest modes responsible for the
breakdown of Hydro+ are $N_{m\T{i}}$ and $N_{\T{i}\T{j}}$ (normalized Wigner
functions obeying Eqs.~(\ref{eq:LN_si}) and~(\ref{eq:LN_ij})). Therefore, within
our formalism we can extend Hydro+ beyond its limit at $k\sim
\xi^{-2}$. In Section~\ref{sec:hydro++} we shall describe how to do
that. Prior to that, in Section~\ref{sec:hydro+}, we shall verify that
in the regime where Hydro+ is applicable, it is in agreement with our
more general formalism.

\subsection{Connection to Hydro+}
\label{sec:hydro+}

The main ingredient of Hydro+ is the entropy density $s_{(+)}$ of the
system in partial equilibrium state where a non-hydrodynamically slow
variable $\varphi$, or more generally, a set of variables $\varphi_q$
indexed by a discrete or continuous index $q$ is not equal to the
equilibrium value $\varphi^{(0)}_q(\eps,n)$ for given $\eps$ and
$n$. For brevity of notations we shall denote such a set of variables
by a bold letter, similar to a vector with components $\phi_q$:
\begin{equation}
  \label{eq:phiq}
  \bm\varphi \equiv \{\varphi_q\}\,.
\end{equation}
The equations of
motion for $\bm\varphi$ describe
relaxation to equilibrium (maximum of $s_{(+)}$) accompanied, in general, by
dilution due to expansion:
\begin{equation}\label{eq:udphi}
   (u\cdot\partial) \bm\varphi = -\bm F_\varphi - \bm A_\varphi\divu,
 \end{equation}
 Second law of thermodynamics requires
 $(F_\varphi)_q=\sum_{q'}\gamma_{qq'}\pi_{q'}$ with
 semi-positive-definite $\gamma$ where $\pi_q$ is the thermodynamic
 ``force'' defined, as usual, via
\begin{equation}
  \label{eq:ds+}
  ds_{(+)}  = \beta_{(+)}d\eps - \alpha_{(+)}dn - \bm\pi\cdot d\bm\varphi.
\end{equation}
where $\bm\pi\cdot\bm\varphi = \sum_q\pi_q\varphi_q$.
The coefficient $A_\varphi$
in Eq.~(\ref{eq:udphi}) describes the response of the variable
$\varphi$ to the expansion or compression of the fluid (since
$\theta=\partial\cdot u$ is the expansion rate). \footnote{For comparison, we
can also cast evolution of hydrodynamic variables or, in general, any
function of $\beps$ and $n$, in the form of Eq.~(\ref{eq:udphi}). In
this case $F_\varphi=0$ and $A_\varphi=\varphi\dot\varphi$. For example,
for charge density $n$: $A_n=n$, since $\dot n=1$, -- the density changes
proportionally with inverse volume, while for the ratio $m=s/n$,
$A_m=0$, since $\dot m=0$.}

The hydrodynamic variables $\eps$ and $u$ obey, as usual, equations of
the energy-momentum conservation. The equation of state enters into
constitutive equations
\begin{equation}
  \label{eq:constitutive+}
  T^{\mu\nu} = \eps u^\mu u^\nu + p_{(+)} \Delta^{\mu\nu} + \Pi^{\mu\nu}
\end{equation}
via
pressure $p_{(+)}$ which, as a function of $\eps$, $n$, and $\phi$, is given by
the Legendre transform of $s_{(+)}$:
\begin{equation}\label{eq:p+}
   {\beta_{(+)}}p_{(+)}={s_{(+)}-\beta_{(+)}\eps+\alpha_{(+)}n+
     \bm\pi\cdot \bm A_\varphi}\,.
 \end{equation}
 This relationship between pressure and entropy is dictated by the
 second law of thermodynamics \cite{Stephanov:2018hydro+}.

Near equilibrium, the deviation of the entropy $s_{(+)}(\eps,n,\varphi)$ from the
equilibrium value $s(\eps,n)$ is quadratic in $\bm\pi$, since
entropy is maximized in equilibrium. The deviation of pressure
$p_{(+)}$ from equilibrium $p$ is linear in $\pi$,
\begin{equation}
  \label{eq:p+ppi}
  p_{(+)} = p + \bm p_\pi\cdot \bm \pi + \mathcal O(\pi^2)\,.
\end{equation}
The coefficient $p_\pi$ can be expressed (see Appendix B in Ref.~\cite{Stephanov:2018hydro+}) using Eqs.~(\ref{eq:ds+})
and~(\ref{eq:p+}), in
terms of the equilibrium value of $\phi$ at given $\eps$ and $n$,
which we denote by $\phieq(\eps,n)$, as
\begin{equation}
  \label{eq:ppi-phieq}
  \beta \bm p_\pi =- w\left(\frac{\partial \bmphieq }{\partial\eps}\right)_n
  -n\left(\frac{\partial \bmphieq }{\partial n}\right)_\eps
  + \bm A_\varphi
  =
  - s\left(\frac{\partial \bmphieq }{\partial s}\right)_m
  + \bm A_\varphi
  = -\bmphieq(\bmphieq)\!\dot{\phantom{I}} + \bm A_\varphi.
\end{equation}

We wish to show that the constitutive equations in Hydro+ with
generalized pressure $p_{(+)}$ are in agreement with the equations we
derived by expanding to quadratic order in fluctuations,
such as Eq.~(\ref{eq:constitutive-R-TJ}).

Application of the Hydro+ approach near the critical point consists of
considering the two-point correlation function of the slowest mode
($m\equiv s/n$): $\varphi\sim\langle\delta m\delta
m\rangle$. Essentially, using our notations
\begin{equation}
  \label{eq:phiN}
  \varphi_q(x) = N_{mm}(x,q)\,. 
\end{equation}
Due to the reparametrization invariance of Hydro+
(see Appendix~C in Ref.~\cite{Stephanov:2018hydro+}), either choice,
$N_{mm}$ or $W_{mm}$, different by a
normalization factor in Eq.(\ref{eq:NW}), will lead to the
same result. The choice of $N_{mm}$ is convenient because in
this case the compression coefficient vanishes: $A_\varphi=0$ (see Eq.~(\ref{eq:LN_ss})).

In order to find non-equilibrium correction to Hydro+ pressure in Eq.~(\ref{eq:p+ppi})  we
need to use the expression for the non-equilibrium contribution to
entropy derived in Ref.~\cite{Stephanov:2018hydro+}
\begin{equation}
  \label{eq:s+-s}
  s^{\rm (neq)} \equiv s_{(+)} - s
  =\frac{1}{2}\int_q\left(
    \log\frac{N_{mm}}{N^{(0)}_{mm}}
    -\frac{N_{mm}}{N^{(0)}_{mm}}+1\right)
\end{equation}
to determine $\pi$:
\begin{equation}
  \label{eq:pi-phi}
  \pi_q \equiv -\frac{\partial s_{(+)}}{\partial\varphi_q}
   = \frac12\left(\frac1{N^{(0)}_{mm}} -
    \frac1{N_{mm}}\right)
  = \frac12{\left(N^{(0)}_{mm}\right)^{-2}}{N_{mm}^{\rm(neq)}}
  + \mathcal O(N_{mm}^{\rm(neq)})^2\,,
\end{equation}  
where
\begin{equation}
N_{mm}^{\rm(neq)}\equiv N_{mm} -
N^{(0)}_{mm}.\label{eq:Nmm-neq}
\end{equation}
The equilibrium value
$N^{(0)}_{mm}$ of $N_{mm}$ also determines the
value of $p_\pi$ via equation~(\ref{eq:ppi-phieq}) with
$\phieq$ replaced by $ N^{(0)}_{mm}$ and $A_\varphi=0$.
Putting this together we find, to linear order in $N_{mm}^{\rm(neq)}$,
\begin{equation}
  \label{eq:pneq}
  p^{\rm (neq)} \equiv  p_{(+)} - p 
  = -\frac T2 \int_q{\left(N^{(0)}_{mm}\right)^{-1}}
  \dot N^{(0)}_{mm}
  {N_{mm}^{\rm(neq)}} = \frac{nT}{2c_p}(1-\dot c_p) \int_q {N_{mm}^{\rm(neq)}} =
  \frac{1-\dot c_p}{2c_pT} G_{mm}^{\rm(neq)} 
\end{equation}
where we used $N_{mm}^{(0)}=c_p/n$, which follows from
Eq.~(\ref{eq:W0}) and
(\ref{eq:NW}) (and can be seen in Eq.~(\ref{eq:LN_ss})) together with the property of the 
log-derivative, Eq.~(\ref{eq:XYdot}). 

We should compare this to the nonequilibrium contribution to pressure
from $N_{mm}$ (which is dominant near critical point due to being
proportional to $c_p$) in Section~\ref{sec:renormalization}:
\begin{equation}
  \label{eq:p-neq}
  p^{\rm (neq)} =  
\left(\frac{\partial p}{\partial\beps}\right)_n (\delta_R\beps-\delta_R^{(0)}\beps)
+ \left(\frac{\partial p}{\partial n}\right)_\beps (\delta_Rn-\delta_R^{(0)}n)
 =  \frac{1-\dot c_p}{2c_pT}G^{\rm (neq)}_{mm}\,,
\end{equation}
which is similar to equilibrium contribution (renormalization of
static pressure) found in Eq.~(\ref{eq:p-renormalized-Lambda}) with
index `$(0)$' replaced by `(neq)'. 
One can see that Hydro+ reproduces these nonequilibrium
 contributions exactly.
We emphasize that this is a very nontrivial
cross-check, involving an elaborate thermodynamic identity for third
derivatives of entropy in Eq.~(\ref{eq:1-cpdot}). This is not
unexpected since Hydro+ formalism 
emerged in Ref.~\cite{Stephanov:2018hydro+} via
very different route, starting from the derivation of the nonequilibrium entropy
functional $s^{(+)}$ in Eq.~(\ref{eq:s+-s}).

\subsection{Hydro++}
\label{sec:hydro++}

Since, as we already discussed above, the fluctuation contributions
are dominated by the modes near the cutoff $\Lambda$, and for critical
fluctuations the role of this cutoff is played by $\xi^{-1}$, the
contributions responsible for the breakdown of ordinary
hydrodynamics 
and of Hydro+ 
are dominated by fluctuations at scale $q\sim\xi^{-1}$. These modes themselves
cannot be described by ordinary hydrodynamics. The dynamics of these
modes is essentially nonlinear and nonlocal (often referred to as
mode-coupling phenomenon). However, this dynamics is universal in the
sense of universality of dynamical critical phenomena and is described
by model H in the classification of Ref.~\cite{Hohenberg:1977}. We
shall, therefore, use the known results from this universality class to describe
the dynamics of these fluctuation modes.

Near the critical point, where the correlation length $\xi$ greatly
exceeds all other {\em microscopic} scales, the description simplifies
due to (static and dynamic) scaling. That means the relaxation
rates, even though no longer polynomial in $q$, as in the hydrodynamic
regime where gradient expansion applies, depend on the $q$ and $\xi$
via functions of only the dimensionless combination $q\xi$ (times a
power of $\xi$). Furthermore, these functions (and the powers of
$\xi$) are universal, i.e., independent of the microscopic composition
or properties of the system close to the critical point in a given
universality class. The universality class relevant for our discussion
is that of model H, defined in Ref.~\cite{Hohenberg:1977} as
{\em dynamic} universality class of liquid-gas phase transitions.

As we already said, the fluctuation kinetic equations, such as
(\ref{eq:LN}), do not apply in the regime $q\xi\sim1$ as they
are. However, a modification of these equations, to match the known
results from model H is possible and shall be described below. We must
emphasize, that unlike the formalism derived in the preceding
sections, which was exact to a certain order in a systematic
expansion, here our out goal is to provide the formalism which
reproduces the physics of critical point fluctuations correctly, but
not necessarily exactly. For once, the exact description would at a
minimum require exact solution to model H, which is not
available. Our approximation is essentially equivalent to a one-loop
approximation introduced by Kawasaki in Ref.~\cite{PhysRevA.1.1750},
which is known to be in good quantitative agreement with experimental
data~\cite{Hohenberg:1977}. 
Similarly to Hydro+ formalism, the purpose of the new extended
formalism, which we shall refer to as Hydro++ in this paper, is to
provide a practical way of simulating the dynamics near the critical
point, e.g., in heavy-ion collisions.

There are two main modifications required.
First of all, we need to
modify equation for $N_{mm}$ to make sure that the
equilibrium correlation function has finite correlation length
$\xi$, i.e., $N_{mm}^{(0)}(x,q)$ must depend on momentum
$q$. We shall express this as
\begin{equation}
  \label{eq:Nss0-cp(q)}
  N_{mm}^{(0)} = \frac{c_p(q)}{n}
\end{equation}
where we defined function $c_p(q)$ in such a way that $c_p(0)=c_p$ is
the usual thermodynamic quantity (heat capacity at constant pressure).
 In this work we shall adopt a simple approximation for the momentum dependence:
\begin{equation}\label{eq:cp-sub}
  c_p\to c_p(q) =  \frac{c_p}{1+(q\xi)^2},
\end{equation}
This is known as Ornstein-Zernike form and is consistent with other
approximations we are making.\footnote{Such as $c_p\sim\xi^2$ instead
  of $c_p\sim\xi^{2-\eta_x}$.} A more sophisticated form and a better
approximation to the exact correlation function (which is not known
exactly as of this writing\footnote{It is the correlation function of
  the 3d Ising model.}) can be used if necessary, see
Ref.~\cite{Stephanov:2018hydro+}.

The second essential modification is required to correctly describe
relaxation rate of the slowest non-hydrodynamic mode, $N_{mm}$. The
critical contribution, $\sim\xi^{-1}$ dominates near the critical
point. It is given  in terms of the Kawasaki function
\begin{equation}
  \label{eq:Kawasaki}
 K(x) =  \frac{3}{4x^2}\left[1+x^2+\left(x^3-x^{-1}\right)\arctan
   x\right] = 1+ \mathcal O(x^2)
\end{equation}
Keeping also noncritical contribution, we can write for the
$q$-dependent rate
\begin{equation}
  \label{eq:Gamma(q)}
  \Gamma(q) \equiv 2\gamma_\lambda(q) q^2 =
2 
\left(\frac{\kappa_0}{c_p} 
+ \frac{T}{6\pi\eta\xi}K(q\xi)\right) q^2
\end{equation}
Note that at small $q$, i.e., $q\xi\ll1$, the rate is given by twice
the diffusion rate $\gamma_\lambda q^2$, where
$\gamma_\lambda=\kappa/c_p$ with $\kappa$ being the zero-frequency
heat conductivity:
\begin{equation}
  \label{eq:lambda0}
  \kappa = \kappa_0 + \frac{c_pT}{6\pi\eta\xi}\,.
\end{equation}
It contains a noncritical contribution $\kappa_0$, but is dominated,
near the critical point, by the critical contribution due to the
fluctuations. The latter increases with $\xi$ as $\kappa\sim\xi$ (in
Kawasaki approximation).

With these two modifications, the equations for Hydro++ we propose read:
\begin{subequations}\label{eq:N_mod}
\begin{align}
\mathcal{L}[N_{mm}] =&
  -2\gl(q) q^2
    \left(     
    N_{mm}-\frac{c_p(q)}{n}           
    \right) 
    -\frac{n}{w}t^{(i)}\cdot\partial_\perp m
    \left(N_{\T{i}m}+N_{m\T{i}}\right),
    \label{eq:Nss_mod}\\
  \mathcal{L}[N_{m\T{i}}] =
 &
   -\left(\geta+     
   \gl(q)\right) q^2N_{m\T{i}} \nonumber\\
 &
   -\partial^\nu u^\mu t_\mu^{(i)} t_\nu^{(j)} N_{m\T{j}}
   -\frac{n}{w}t^{(j)}\cdot\partial_\perp m N_{\T{j}\T{i}}
   +Tn\left(        
   \frac{1}{c_p(q)}t^{(i)}\cdot\partial_\perp m+\frac{T}{w}t^{(i)}\cdot\partial_\perp\alpha\right)                                               
    N_{mm} 
    ,\label{eq:Nsi_mod}\\
  \mathcal{L}[N_{\T{i}\T{j}}] =&
  -2\geta q^2\left( N_{\T{i}\T{j}}-\frac{Tw}{n}\delta_{ij} \right) \nonumber\\ &
    -\partial^\nu u^\mu
    \left(
    t_\mu^{(i)} t_\nu^{(k)} N_{\T{k}\T{j}}+t_\mu^{(j)} t_\nu^{(k)} N_{\T{i}\T{k}}
    \right)
    +\frac{\alpha_pT^2n}{w}
    \partial_\perp^\mu p\left(
    t^{(i)}_\mu N_{m\T{j}}
    + t^{(j)}_\mu N_{\T{i}m}
  \right),
  \label{eq:Nij_mod}
\end{align}
\end{subequations}
where again, $\alpha_p=(1-\dot T/c_s^2)/Tn$. The function $\gamma_\lambda(q)$ is defined in
Eq.~(\ref{eq:Gamma(q)}).  The presence of function $c_p(q)$, defined in
Eq.~(\ref{eq:cp-sub}), in
Eq.~(\ref{eq:Nsi_mod}) ensures important property of $N_{m\T{i}}$
in equilibrium -- proportionality to $\partial\alpha$, which follows
from the second law of thermodynamics as we already discussed in
connection with Eq.~(\ref{eq:LN_si}).~\footnote{Heuristically, one can
  obtain Eq.~(\ref{eq:Nsi_mod}) from Eq.~(\ref{eq:LN_si}) by
  preforming substitution of $c_p$ according to Eq.~(\ref{eq:cp-sub}) } Other terms may also contain ``formfactors'',
i.e., functions of $q\xi$, which could be determined from a more
detailed calculation of 3-point functions in model-H. We leave such
and similar refinements to future work. It is likely that given the
general degree of applicability of hydrodynamics in heavy-ion
collisions these will be beyond the experimentally relevant precision.

Eq.~(\ref{eq:Nss_mod}) describes relaxation of the slowest
nonhydrodynamic mode, $N_{mm}$, to equilibrium given by
Eq.~(\ref{eq:Nss0-cp(q)}). It would be identical to the corresponding
Hydro+ equation in Ref.~\cite{Stephanov:2018hydro+}, but for the last
term describing the coupling to next-to-slowest mode, $N_{m\T{i}}$.
Because $c_\p\sim \xi^2$ diverges at the critical point, this term is
indeed much smaller than the first term sufficiently close to the
critical point. However, if we want to interpolate Hydro+ description
close to the critical point with dynamics of fluctuations away from
the critical point this term has to be kept.

\subsection{Conductivity and its frequency dependence in Hydro++}
\label{sec:cond-its-freq}

Let us discuss physics described by Eqs.~(\ref{eq:N_mod}) which is
pertinent to the breakdown of Hydro+ and its crossover to Hydro++.

We can use Eqs.~(\ref{eq:N_mod}) to determine the critical contribution
$\lambda_\xi$ to the conductivity $\lambda$ and verify it diverges as
$\xi\to\infty$. Following the procedure of renormalization described in
Section~\ref{sec:renormalization} we now find that
$W^{(1)}_{m\mu}$, i.e., the part of $W_{m\mu}$ linear in
gradients, is given by Eq.~(\ref{eq:W1}) with a simple substitution
$c_p\to c_p(q)$.
This, in turn, makes the integral of $W_{m\mu}^{(1)}$,
$G^{(1)}_{m\mu}(x)$, finite. The cutoff is now essentially given by
$1/\xi$, instead of $\Lambda$.  This means that instead of
Eq.~(\ref{eq:G(1)}) we find, using $c_p(q)$ in Eq.~(\ref{eq:cp-sub}),  $\Lambda\to \pi/(2\xi)$.
Substituting this result into  equation~(\ref{eq:lambda_R}) for the renormalized
conductivity we find a contribution to renormalized conductivity which
diverges with~$\xi$:
\begin{equation}
  \label{eq:lambda-xi}
\lambda_\xi =
\left(\frac{Tn}{w}\right)^2\frac{c_pT}{6\pi\eta\xi}\sim \xi^1
\end{equation}
-- a well-known result \cite{Hohenberg:1977}.
We used the fact
that $c_p\sim\xi^2$. In particular, since
$\gl\sim\lambda/c_p\sim \xi^{-1}$ we neglected $\gl$ compared to $\geta\sim\xi^0$.
Denoting the noncritical contribution to conductivity by $\lambda_0$
we can write the total physical conductivity as
\begin{equation}
  \label{eq:lambda-total}
  \lambda = \lambda_0 + \lambda_\xi = \lambda_0 +
  \left(\frac{Tn}{w}\right)^2\frac{c_pT}{6\pi\eta\xi} 
= \left(\frac{Tn}{w}\right)^2
\left(\kappa_0
+ \frac{c_pT}{6\pi\eta\xi}\right)
\end{equation}
Note that the relaxation rate $\Gamma(q)$ in Eq.~(\ref{eq:Gamma(q)})
at $q=0$  matches twice the relaxation rate of the diffusive mode,
$\gamma_\lambda=\kappa/c_pq^2$, as it should since this is the
relaxation rate of the corresponding {\em two}-point function.

In the Hydro+ formulation in Ref.~\cite{Stephanov:2018hydro+} the value
of conductivity was given directly by Eq.~(\ref{eq:lambda-xi}). In our
more general approach, which we refer to as Hydro++, the divergent
value of the conductivity is generated ``dynamically'' via the
contribution  of the fluctuation mode $W_{m\T{i}}$ (via $\N^{m\mu}(x)$) to
the constitutive equation for the current in Eq.~(\ref{eq:avJmu}). The
value of $\lambda=\lambda_0$ in Eq.~(\ref{eq:diffusive-nu}) is finite
as $\xi\to\infty$. This is similar to the way divergence of bulk
viscosity with $\xi\to\infty$ is generated in Hydro+ (and, by extension,
also in Hydro++), see Ref.~\cite{Stephanov:2018hydro+}. Similarly to Hydro+,
which describes frequency dependence of bulk viscosity (and sound speed)
Hydro++ describes the frequency dependence of the kinetic coefficient
$\lambda$. We shall consider it below.

Hydro++ allows us to see how Hydro+ breaks down when $k$ (or, more
precisely, the sound frequency $\omega=k/c_s$ at this wave number) exceeds a
value of order $\xi^{-2}$. This happens because the characteristic 
relaxation rate of
the mode
$W^{(1)}_{mi}$ responsible for $\lambda_\xi$ contribution also
vanishes as $\xi\to\infty$:
\begin{equation}
\Gamma_\xi'\equiv \geta q^2\Big|_{q\xi=1}\sim
\xi^{-2}.\label{eq:Gamma_xi'}
\end{equation}
This is next-to-slowest relaxation rate, after the
characteristic relaxation rate of $W_{mm}$, given by\footnote{More
  precisely, $\Gamma_\xi\sim\xi^{x_\lambda-4+\eta_x}=\xi^{-z}$ and
  $\Gamma_\xi'\sim \xi^{x_\eta-2}=\xi^{z+d-8}$ (see also footnote~\ref{fn:z}).}
\footnote{Since the
  bulk viscosity is proportional to the longest microscopic relaxation
  time, vanishing $\Gamma_\xi$ is responsible for the divergence of the
  the bulk viscosity $\zeta\sim c_s^2/\Gamma_\xi\sim
  \xi^{z-\alpha/\nu}$. In the Kawasaki approximation $\zeta\sim\xi^3$.
Since $\zeta$ is the coefficient of the gradient expansion, the
expansion breaks down at $k\xi^3\sim1$, which is an alternative way to
see that ordinary hydrodynamics breaks down at this scale.}
\begin{equation}
\Gamma_\xi\equiv 2\gl q^2|_{q\xi=1}\sim\xi^{-3}\,.\label{eq:Gamma_xi}
\end{equation}

\begin{figure}[ht]
  \centering
  \includegraphics[height=18em]{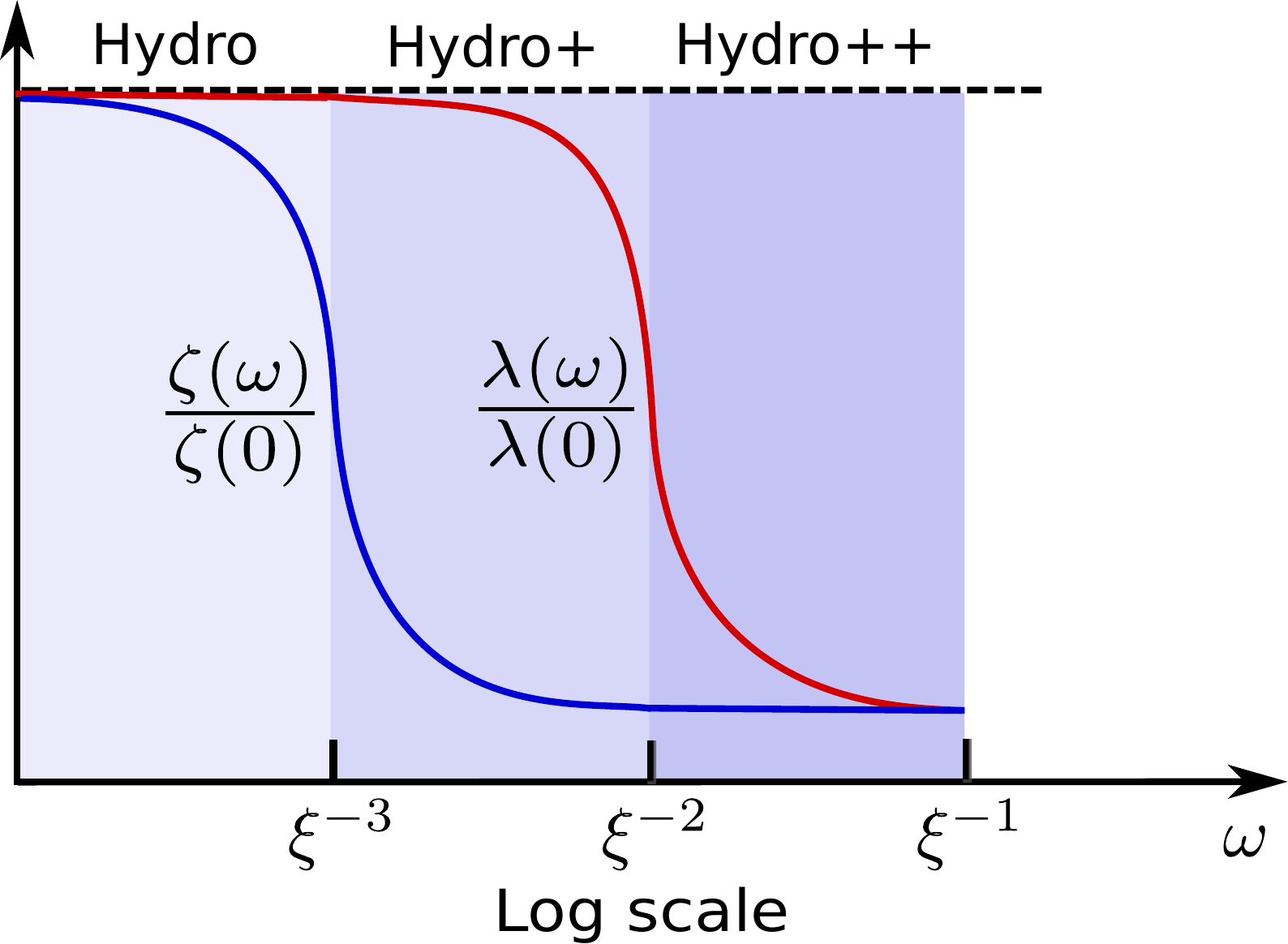}
  \caption{Frequency dependence of transport coefficients
    $\zeta(\omega)$ and $\lambda(\omega)$ in the vicinity of a
    critical point, where the divergence of $\xi$ leads to several
    distinct regimes characterized by frequency $\omega$ (or
    corresponding wave-number $k=\omega/c_s$).  The crossover from
    ordinary hydrodynamics (Hydro) to Hydro+ is marked by the fall-off
    of $\zeta(\omega)$ at $\omega\sim\Gamma_\xi\sim\xi^{-3}$, while
    the Hydro+ itself breaks down at
    $\omega\sim\Gamma_\xi'\sim\xi^{-2}$ as signaled by the fall-off of
    $\lambda(\omega)$, when the crossover to Hydro++ regime occurs. Of
    course, in ordinary hydrodynamics both transport coefficients are constants
    independent of frequency (dashed line), while in Hydro+, which does
    describe the fall-off of $\zeta(\omega)$, the coefficient
    $\lambda$ is still a constant. Hydro++ describes the fall-off of
    both $\zeta(\omega)$ and $\lambda(\omega)$.}
  \label{fig:zeta-lambda}
\end{figure}

As discussed in Ref.~\cite{Stephanov:2018hydro+}, when the evolution
rate (or sound frequency) $\omega$ exceeds $\Gamma_\xi$ the mode
$W_{mm}$ is no longer able to relax to its equilibrium value which
is responsible for the divergence of the bulk viscosity. Therefore,
the divergent contribution to the bulk viscosity is ``switched off'' for
$\omega>\Gamma_\xi$.  Similarly, when the evolution rate (or sound
frequency) $\omega$ exceeds $\Gamma_\xi'$, the next to slowest mode,
$W_{m\mu}$, is no longer able to relax to its zero-frequency value
given in Eq.~(\ref{eq:W1}). As a result, the contribution of
$W_{m\mu}$ to the current in Eq.~(\ref{eq:avJmu}) ``switches
off''. This behavior and corresponding scales are illustrated in Fig.~\ref{fig:zeta-lambda}.

We can further quantify this description by considering the dependence
of $W^{(1)}_{m\mu}$ on frequency
following the same procedure as in Section~\ref{sec:long-time-tails}.
Combining the substitution in
Eq.~(\ref{eq:Womega-q2}) with the substitution (\ref{eq:cp-sub}) in
Eqs.~\eqref{eq:wt-Jmu} and~\eqref{eq:W1} we
find for frequency-dependent leading critical contribution to conductivity:
\begin{equation}
  \label{eq:lambda-xi-omega}
  \lambda_\xi(\omega) = \lambda_\xi(0)
  F_\lambda\left({\omega}/{\Gamma_\xi'}\right)\,,
\end{equation}
where
\begin{equation}\label{eq:F-lambda}
F_\lambda(y)
=\frac2\pi \int_0^\infty \frac{dx x^2}{(x^2-iy)(1+x^2)}
=\frac{1}{1+\sqrt{y/i}}\,.
\end{equation}
We do not need to regularize and subtract a divergence, as we did in
Section~\ref{sec:long-time-tails}, because the divergence is tamed by
the fall-off of $c_p(q)$ at large $q$.

At small $\omega\ll\Gamma_\xi'$ Eq.~(\ref{eq:lambda-xi-omega})
reproduces the power-law non-analytic dependence characteristic of the
long-time tails in Eq.~(\ref{eq:lambda_omega}):
$\lambda_\xi(\omega)-\lambda_\xi(0)\sim\lambda_\xi(0)\omega^{1/2}$. Not
surprisingly, since, compared to Section~\ref{sec:long-time-tails}, we
only changed the nature of the cutoff $\Lambda$. At large $\omega$ we
find $\lambda_\xi(\omega)\sim \omega^{-1/2}$ with no $\xi$ dependence
as expected from scaling behavior characterizing this regime.
\footnote{As before (see footnote~\ref{fn:z}), the exact value of the
  scaling exponent in $\lambda_\xi(\omega)\sim\omega^{-1/2}$ differs slightly
  from the rational value~$-1/2$. The exact value in model H following
  from dynamic scaling $-x_\lambda/(2-x_\eta)={-(4-\eta-z)/(d+2-z)}$
  is approximately~$-1/2$ in the Kawasaki approximation we are using,
  which corresponds to $z\approx 3$ and $\eta_x\approx 0$.}
The dependence of $\lambda_\xi$ on $\omega$ described by
Eqs.~(\ref{eq:lambda-xi-omega}) and~(\ref{eq:F-lambda}) corresponds to
the physics we anticipated -- the large critical contribution
``switches off'' when $\omega\gtrsim\Gamma_\xi'$.

It may also be helpful to note that while real part of
$\lambda_\xi(\omega)$ corresponds to (frequency dependent)
conductivity, its imaginary part (divided by $\omega$) is the electric
permittivity. 

One can also understand frequency dependence as a time-delayed
medium response to gradient of density, i.e., $\partial\alpha$. The
diffusive current induced by the gradient is given by 
\begin{equation}
  \label{eq:Jalpha-t}
  J_\xi(t) =  \lambda_\xi \int_{-\infty}^{t} dt' \Gamma_\xi' 
  \tilde F_\lambda\left(\Gamma_\xi'      
(t-t')\right) \partial\alpha(t')\,. 
\end{equation}
The delay is given by the Fourier transform of
$F_\lambda(y)$:
\begin{equation}
  \label{eq:tildeF}
  \tilde F_\lambda(\tilde y) = \sqrt{\frac{1}{\pi \tilde y}}
  -e^{\tilde  y}\text{erfc}\left(\sqrt{\tilde y}\right)\,.
\end{equation}
As a function of $t-t'$ it has a characteristic width given by
$1/\Gamma_\xi'\sim\xi^{2}$ and becomes $\delta$-function
in the limit $\xi\to0$ corresponding to instantaneous response. At
large $t-t'$ it falls off as $(t-t')^{-3/2}$ typical of the long-time
hydrodynamic tails.

The discussion of the frequency dependence of conductivity here
carries many similarities to the discussion of the bulk viscosity in
Ref.~\cite{Stephanov:2018hydro+}. For completeness, let us present the
calculation of the leading critical contribution to the bulk viscosity
in Hydro++, which, of course, gives the same result as Hydro+. Near
the critical point the leading contribution of fluctuations to the
bulk viscosity comes from $G_{mm}^{(1)}$ in Eq.~(\ref{eq:p-neq}). 
In Hydro++ the corresponding $W_{mm}^{(1)}$ is given in
Eq.~(\ref{eq:N1pmij}), with the substitution of $c_p$ with $c_p(q)$
as in Eq.~(\ref{eq:cp-sub}), as well as $\gamma_\lambda$ with
$\gamma_\lambda(q)$ according to Eq.~(\ref{eq:Gamma(q)}). As a result
we obtain for the leading critical contribution to bulk viscosity:
\begin{equation}
  \label{eq:zeta(omega)}
  \zeta_\xi(\omega) = \frac{3}{\pi} \eta\, \dot\xi^2
  F_\zeta \left(\frac{\omega}{\Gamma_\xi}\right),
\end{equation}
where we used $\dot c_p = 2\dot \xi$ (according to scaling
$c_p\sim\xi^2$) and $\Gamma_\xi=T/(3\pi\eta\xi^3)$ (according to
Eqs.~(\ref{eq:Gamma_xi}) and~(\ref{eq:lambda-xi})).  
We introduced
\begin{equation}
  \label{eq:F-zeta}
  F_\zeta(y) = \int_0^\infty 
\frac{dx x^2}{(x^2K(x)-iy)(1+x^2)^2}\,.
\end{equation}
This is a known result in Kawasaki
approximation~\cite{PhysRevA.1.1750,Stephanov:2018hydro+}.\footnote{As
  we already discussed, Kawasaki approximation only gives a good
  approximation to the correct scaling behavior. To match the exact
    scaling behavior on would need a more elaborate choice of the
  substitution in Eq.~(\ref{eq:cp-sub}), see e.g.,
  Refs.\cite{PhysRevA.1.1750,PhysRevE.55.403,Stephanov:2018hydro+}.}  At $\omega=0$
Eq.~(\ref{eq:zeta(omega)}) gives $\zeta_\xi(0)\sim\xi^3$ (according to
the scaling of $\dot\xi\sim\xi^{3/2}$). This large critical contribution is
``switched off'' via function $F_\zeta$ when
$\omega>\Gamma_\xi$.\footnote{
The large $\omega$ asymptotics $\zeta_\xi(\omega)\sim\omega^{-1}$ in Kawasaki approximation
is close to the exact asymptotics $\omega^{-1+\alpha/(z\nu)}$. 
}

 The
resulting behavior is illustrated in Fig.~\ref{fig:zeta-lambda}
together with the behavior of $\lambda(\omega)$.


\section{Discussion,  conclusions and outlook}
\label{sec:conclusions}
In this paper we continued the development of the
deterministic approach to fluctuation hydrodynamics for an arbitrary
relativistic flow.  We extended the method developed in
Ref.~\cite{An:2019rhf} to the case of a fluid carrying conserved
charge. In QCD the relevant charge is the baryon number.  Our
ultimate goal is practical -- a formalism which would allow to
simulate heavy-ion collisions with dynamical effects of fluctuations,
especially relevant for the QCD critical point search. 

We would like to emphasize that, despite its practical aim, this
formalism is based on a systematic and {\em controllable} expansion, similar
to the effective field theory formalism in quantum field theory. The
expansion parameter in hydrodynamics is the ratio of the wave-number
$k=1/L$ associated with background flow and density gradients to a
microscopic scale which sets the scale of hydrodynamic coefficients
and which we denote $1/\lmic$. This
allows us to view hydrodynamics as an effective theory.

  Instead of directly solving stochastic hydrodynamic equations, we
  convert them into a hierarchy of equations for equal-time
  correlation functions, which we truncate at two-point
  correlators. This truncation is controlled by the same expansion
  parameter as the gradient expansion in hydrodynamics. One can see
  how the relevant power counting emerges by considering the effects
  of fluctuations on the constitutive equations for stress tensor (or
  conserved current). In stochastic hydrodynamics the noise is local,
  i.e., it is only correlated inside a hydrodynamic cell, as reflected
  in the $\delta$-function value of the two-point noise correlator in
  Eq.~(\ref{eq:fdt}). This locality is the source of short-distance
  singularities, similar to ultraviolet singularities in quantum field
  theories. Hydrodynamics is regulated by finiteness of the cell size,
  which we denote by $b\gg\lmic$, equivalent to wave-number cutoff
  $\Lambda=1/b$. As a function of this regulator, the square variance
  of the noise in each cell is proportional to $\Lambda^3$ -- the regulated
  value of the $\delta$-function. This is, of course, the source of
  the cutoff dependent contribution to renormalized pressure in
  Eq.~(\ref{eq:p-renormalized-Lambda}) and, as such, is not of
  physical relevance.

The physically consequential contribution comes from the fluctuations
whose relaxation time is comparable to the evolution time of the
background.  Correspondingly, this scale, characterized by wave-number
$q_*$, can be estimated by the condition $\gamma q_*^2\sim c_s k$.
The effect of these fluctuations is the delayed or nonlocal response
to perturbations of the background (such as long-time tails) and
cannot be simply absorbed by renormalization of the local hydrodynamic
parameters such as pressure or transport coefficients. Since
$q_*\ll \Lambda$, the noise on these longer distance scales,
$\ell_*=1/q_*$, averages out and the magnitude of the fluctuations is
effectively reduced by a factor $(b^3/\ell_*^3)^{1/2}=(q_*/\Lambda)^{3/2}$ -- the inverse
of the square root of the number of uncorrelated cells in a region of
linear size $\ell_*$ -- the familiar random walk factor. Therefore the
physically relevant magnitude of the fluctuations, obtained by
averaging over scales $\ell_*$ is given by
$\Lambda^{3/2}\times (q_*/\Lambda)^{3/2}\sim q_*^{3/2}\sim
k^{3/4}$. It is cutoff independent, of course. Therefore, the
two-point correlator of these fluctuations contributes at order
$k^{3/2}$, suppressed compared to first-order gradients, but more
important than second order gradients. Similarly, the contribution of
$n$-point functions, due to higher order nonlinearities in the
constitutive equations, would come at order $k^{3n/4}$
One can see
that the hierarchy of higher-point contributions is controlled by a
power of~$k$, or more precisely, a power of dimensionless parameter
$k\lmic=\lmic/L\ll1$. 

The equations we derive form a closed set of deterministic equations
which can be solved numerically. The one-point functions (averaged
values of hydrodynamic variables) obey conservation
equations~(\ref{eq:conservation-TJ}). The constitutive
equations~(\ref{eq:constitutive-R-TJ}) contain contributions
$\widetilde T^{\mu\nu}$ and $\widetilde J^\mu$ which are given in
terms of the subtracted two-point functions $\wt G$ in
Eqs.~(\ref{eq:wt-TJ}). The unsubtracted two-point functions $G$
are evaluated at coinciding points and therefore contain short-range
singularities. When unsubtracted $G$ are expressed in terms of the
wave-number integrals of the Wigner functions Eq.~(\ref{eq:G(x)-W}),
these singularities appear as ultra-violet divergences which need to
be subtracted. The unsubtracted Wigner functions are obtained by
solving equations~(\ref{eq:Npm}) and~(\ref{eq:LN}), rescaling
according to Eqs.~(\ref{eq:Npm-Wpm}) and~(\ref{eq:NW}) and
substituting into the matrix in Eq.~(\ref{eq:WAB-WcAcB}). The
subtraction of terms of zero and first order in gradients, $W^{(0)}$
and $W^{(1)}$, given by Eqs.~(\ref{eq:W0}) and~(\ref{eq:W1})
respectively, can be done analytically, and either before or after
solving equations~(\ref{eq:Npm}) and~(\ref{eq:LN}), depending on
numerical efficiency. The resulting solutions to one-point and
two-point equations will describe evolution of the average
hydrodynamic variables, their fluctuations, as well as the feedback of
the fluctuations on the evolution of average quantities.

With the equations we derived we can now also describe the essential
features of the hydrodynamic evolution near the QCD critical
point. The critical phenomena are originating from the divergence of
the correlation length $\xi$. The phenomenon of the most consequence
for hydrodynamics is the critical slowing down. Since it is caused by
the fluctuations of the slowest diffusive mode out of equilibrium, our
formalism is ideally suited to accommodate and describe this
phenomenon. The formalism of Hydro+ introduced earlier in
Ref.~\cite{Stephanov:2018hydro+} is based on the same observation and
adds the two-point correlation function of the diffusive mode to
hydrodynamics to describe critical slowing down. The approach in the
present paper is very different from the derivation in
Ref.~\cite{Stephanov:2018hydro+}, therefore, the exact agreement
between the results is a nontrivial check on the validity of both
derivations.

Since, our present approach is more general, we can now connect Hydro+
description of critical fluctuations to description of ordinary
fluctuations away from the critical point. Because the validity of
Hydro+ is limited by the relaxation rate of the next-to-slowest mode,
and this mode, absent in Hydro+, is now a part of our description, we are able to extend the
validity of hydrodynamic description closer to the critical point than
Hydro+. We propose a set of equations, which we call Hydro++ which
could accomplish this. It should be kept in mind that, unlike the
systematic approach taken in the rest of the paper, the Hydro++
equations~(\ref{eq:N_mod}) are an attempt to interpolate between the
description of fluctuations outside of the critical regime and the
known properties of the fluctuations in the critical, scaling regime
described by model H (in the standard classification of
Ref.~\cite{Hohenberg:1977}). While the hydrodynamic description still
works for the background gradients for which $k\lmic\sim k\xi\ll1$, it
breaks down for critical {\em fluctuations}, for which $q\xi\sim1$. This
means that the coefficients become non-polynomial in $q$ and that the
theory becomes fully nonlinear and the truncation to two-point
functions is no longer, strictly speaking, controllable. However, it
is known from the studies of model H that the results obtained in one-loop
(Kawasaki) approximation are in  good quantitative
agreement with experiment~\cite{Hohenberg:1977}. Therefore we propose a set of
equations~(\ref{eq:N_mod}) which incorporate the model H physics at
the corresponding level of approximation. This approach is similar to the
one taken in the derivation of Hydro+ and extends the region of
applicability closer to the critical point. More precisely, while
Hydro+ breaks down at $k\sim\xi^{-2}$, the validity of Hydro++ extends
to $k\sim\xi^{-1}$. The physical phenomenon which leads to breakdown
of Hydro+ is the frequency dependence of (i.e., time-lag
of) conductivity, which is described by the next-to-slowest mode in
Hydro++.

Once the fluctuation hydrodynamics in
the deterministic approach is implemented
in a fully functional hydrodynamic code\footnote{A simplified example of such
an implementation for Hydro+ is presented in Ref.\cite{Rajagopal:2019hydro}.}, the extension to full Hydro++
approach should be straightforward and will allow eventual comparison
with heavy-ion collision experiments not only near, but also away from
the critical point. However, additional developments are needed to
make this comparison more impactful. First of all, it should be
straightforward to generalize this approach to multiple conserved
charges. In the case of QCD, of course, fluctuations of isospin are a
primary candidate. We have not included these fluctuations in our
description because they are not exhibiting singularities near the
critical point, unlike the baryon number fluctuations, which lead to
signatures of the QCD critical point~\cite{Hatta:2003wn}.
Furthermore, the approach must be extended to description of
non-Gaussian fluctuations, which are related to most sensitive
signatures of the critical
point~\cite{Stephanov:2008qz,Stephanov:2011pb}. 
This means going beyond two-point
correlators considered in this paper. It would also be interesting and
important for comparison with experiment to consider the extension of
this approach to the fluctuations near the first-order phase
transition, which is, of course, an inseparable part of the physics
near a critical point. We defer these and other pertinent developments to
future work.


\acknowledgements

We thank Mauricio Martinez, Derek Teaney and Yi Yin for
helpful discussions.  This work is supported by the U.S. Department of
Energy, Office of Science, Office of Nuclear Physics, within the
framework of the Beam Energy Scan Theory (BEST) Topical Collaboration
and grant No. DE-FG0201ER41195.

\appendix
\section{Useful thermodynamic relations and derivatives}
\label{sec:derivation}

The thermodynamic derivatives appearing in Eq.~\eqref{eq:dedp-dmdn} and \eqref{eq:dalpha-dmdn}, i.e.,
\begin{equation}
  d\eps=\eps_mdm+\eps_pdp, \quad dn=n_mdm+n_pdp, \quad d\alpha=\alpha_mdm+\alpha_pdp,
\end{equation}
 are defined in the standard notation 
\begin{equation}\label{eq:coeff_def_2nd}
  \eps_m\equiv\left({\partial\eps\over\partial m}\right)_p, \quad \eps_p\equiv\left(\frac{\partial\eps}{\partial p}\right)_m, \quad n_m\equiv\left(\frac{\partial n}{\partial m}\right)_p, \quad n_p\equiv\left(\frac{\partial n}{\partial p}\right)_m,
  \quad \alpha_m\equiv\left({\partial\alpha\over\partial m}\right)_p, \quad \alpha_p\equiv\left({\partial\alpha\over\partial p}\right)_m.\\
\end{equation}
To obtain the relations of the second order thermodynamic coefficients, Eq.~\eqref{eq:2ndordercoeff}, we begin with the thermodynamic relations coming from the first law of thermodynamics (Eq.~\eqref{eq:ds}):
\begin{equation}
  d\eps=Tndm+\frac{w}{n}dn, \quad dp=\frac{w}{T}dT+Tnd\alpha, \quad d\left(\frac{w}{n}\right)=Tdm+\frac{1}{n}dp,
\end{equation}
from which we obtain
\begin{equation}\label{eq:derivative_appendix}
\begin{gathered}
  \left(\frac{\partial\eps}{\partial m}\right)_n=\left(\frac{\partial p}{\partial\alpha}\right)_T=Tn, \quad \left(\frac{\partial\eps}{\partial n}\right)_m=\frac{w}{n}, \quad \left(\frac{\partial n}{\partial m}\right)_\eps=-\frac{Tn^2}{w}, \quad \left(\frac{\partial\alpha}{\partial T}\right)_p=-\frac{w}{T^2n}, \quad \left(\frac{\partial T}{\partial p}\right)_m=-\frac{n_m}{n^2}.
\end{gathered}
\end{equation}
Therefore 
\begin{equation}\label{eq:eps_m}
\begin{aligned}
  \eps_m&=\left(\frac{\partial\eps}{\partial m}\right)_n+\left(\frac{\partial\eps}{\partial n}\right)_m\left(\frac{\partial n}{\partial m}\right)_p=Tn\left(1+\frac{n_mw}{Tn^2}\right)=(Tn)^2\left(\left(\frac{\partial\alpha}{\partial p}\right)_T+\left(\frac{\partial\alpha}{\partial T}\right)_p\left(\frac{\partial T}{\partial p}\right)_m\right)=(Tn)^2\alpha_p\,.
\end{aligned}
\end{equation}
Noting that
\begin{equation}
  \dot T=\frac{n}{T}\left(\frac{\partial T}{\partial n}\right)_m=\frac{n}{T}\left(\frac{\partial\eps}{\partial n}\right)_m\left(\frac{\partial T}{\partial p}\right)_m\left(\frac{\partial p}{\partial\eps}\right)_m=-\frac{c_s^2n_mw}{Tn^2}=c_s^2\left(1-\frac{\eps_m}{Tn}\right)=c_s^2\left(1-\alpha_pTn\right),
\end{equation}
we obtain
\begin{equation}
  \eps_m=Tn\left(1-\frac{\dot T}{c_s^2}\right), \quad  \alpha_p=\frac{1}{Tn}\left(1-\frac{\dot T}{c_s^2}\right),
\end{equation}
demonstrating the first identity in Eq.~\eqref{eq:2ndordercoeff}. Likewise, the remaining nontrivial identities in Eq.~\eqref{eq:2ndordercoeff} are obtained by using Eq.~\eqref{eq:derivative_appendix} and turn out to be
\begin{equation}
\begin{gathered}
  n_m=\left(\frac{\partial n}{\partial m}\right)_\eps+\left(\frac{\partial n}{\partial\eps}\right)_m\left(\frac{\partial\eps}{\partial m}\right)_p=\frac{n}{w}\left(\eps_m-Tn\right)=\frac{Tn^2}{w}\left(Tn\alpha_p-1\right)=-\frac{\dot TTn^2}{c_s^2w},\\
  n_p=\left(\frac{\partial n}{\partial\eps}\right)_m\left(\frac{\partial\eps}{\partial p}\right)_m=\frac{n}{c_s^2w}, \quad \alpha_m=\left(\frac{\partial\alpha}{\partial T}\right)_p\left(\frac{\partial T}{\partial m}\right)_p=-\frac{w}{nT^2}\left(\frac{\partial T}{\partial m}\right)_p=-\frac{w}{c_pT}\,.
\end{gathered}
\end{equation}
Throughout the above derivation, we have used the definition
\begin{equation}
  c_s^2\equiv\left(\frac{\partial p}{\partial\eps}\right)_m, \quad c_p\equiv Tn\left(\frac{\partial m}{\partial T}\right)_p.
\end{equation}

Similarly, the third order derivatives appearing in Eq.~\eqref{eq:epsn_2nd_expansion} are defined by
\begin{equation}\label{eq:coeff_def_3rd}
\begin{gathered}
\eps_{mm}\equiv\left(\frac{\partial^2\eps}{\partial m^2}\right)_p\,,\quad \eps_{pp}\equiv\left(\frac{\partial^2\eps}{\partial p^2}\right)_m\,, \quad n_{mm}\equiv\left(\frac{\partial^2\eps}{\partial m^2}\right)_p\,,\quad n_{pp}\equiv\left(\frac{\partial^2\eps}{\partial p^2}\right)_m.
\end{gathered}
\end{equation}
Note that the mixed third order derivatives are not presented here as
they are not relevant in our calculation. The results presented in
Eq.~\eqref{eq:3rdordercoeff} can be derived straightforwardly from
the known expression of second order thermodynamic derivatives given
above. We leave this exercise to the reader.

\section{Comparison to known results}
\label{sec:comparison}

Our new results can be compared to some existing results in the literature
for several special cases.

A charged fluid studied in Ref.~\cite{Martinez:2018} is (i) conformal
and (ii) undergoes a boost-invariant (Bjorken) expansion. Thermodynamic
functions of a conformal fluid satisfy $\dot T=c_s^2=1/3$,
$\eps_m=\alpha_p=\dot c_s=0$, $\dot c_p=1$ as summarized in
Table~\ref{tab:limits}. The boost-invariant flow implies that
$a_\mu=\omega_{\mu\nu}=0$ and spatial gradients of background scalar
fields vanish (e.g., $\partial_{\perp\mu}\alpha=0$). Under these
conditions our results are significantly simplified. Since in a
boost-invariant Bjorken flow, the charge does not diffuse due to the
absence of background gradients forbidden by boost-invariance, in
order to generate the dissipative (ohmic) charge current, one needs to
apply an external electric field to the system. Adding such a source
term is indispensable for obtaining such important results as
renormalized or frequency-dependent conductivity in
Ref.~\cite{Martinez:2018}.  We find that except for a few minor typos,
our Eqs.~\eqref{eq:eta_R} and \eqref{eq:lambda_R} for renormalized
transport coefficients, as well as Eq.~\eqref{eq:lambda_omega} and
\eqref{eq:viscosities_omega} for frequency-dependent transport
coefficients, reduce to Eq.~(51) and (50) in Ref.~\cite{Martinez:2018}
respectively. Notice that $\zeta=0$ in conformal fluid.

Despite this agreement with 
Ref.~\cite{Martinez:2018}, there are still some mismatches. For
example, our Eq.~\eqref{eq:1st_ord} would have matched Eq.~(62) in
\cite{Martinez:2018} in the absence of source term, if it wasn't for the last
term $\sim1/\tau$ in Eq.~(64a), which should be $\sim(1+c_s^2)/\tau$
according to our results. 

To compare one of the key
results of our paper, Eqs.~\eqref{eq:LN}, to
Ref.~\cite{Martinez:2018} we need to rescale our Wigner functions,
somewhat similarly to Eq.~\eqref{eq:NW}, 
\begin{equation}
  N^{\text{B}}_{mm}\equiv\frac{W_{mm}}{\tau}, \quad N^{\text{B}}_{\T{i}\T{j}}\equiv\frac{W_{\T{i}\T{j}}}{\tau},
\end{equation}
where $\tau$ is the Bjorken proper time coordinate, and express the unit vectors in spherical coordinates
\begin{equation}
  \hat q=(0,\sin\theta\cos\phi,\sin\theta\sin\phi,\cos\theta),\quad t^{(1)}=(0,-\sin\phi,\cos\phi,0),\quad t^{(2)}=(0,\cos\theta\cos\phi,\cos\theta\sin\phi,-\sin\theta).
\end{equation}
Then, our equations read
\begin{eqnarray}
&&\partial_\tau N^{\text{B}}_{mm} = -2\gl q^2\left(N^{\text{B}}_{mm}-\frac{c_pT^2}{\tau}\right)-\frac{2+2\left(1-\alpha_pTn\right)c_s^2}{\tau} N^{\text{B}}_{mm},\nn
&&\partial_\tau N^{\text{B}}_{m\T{1}} = -(\geta+\gl) q^2N^{\text{B}}_{m\T{1}}-\frac{2+\left(1-\alpha_pTn\right)c_s^2}{\tau} N^{\text{B}}_{m\T{1}},\nn
&&\partial_\tau N^{\text{B}}_{m\T{2}} = -(\geta+\gl) q^2N^{\text{B}}_{m\T{2}}-\frac{2+\left(1-\alpha_pTn\right)c_s^2+\sin^2\theta}{\tau} N^{\text{B}}_{m\T{2}},\nn
&&\partial_\tau N^{\text{B}}_{\T{1}\T{1}} = -2\geta q^2\left(N^{\text{B}}_{\T{1}\T{1}}-\frac{Tw}{\tau}\right)-\frac{2}{\tau} N^{\text{B}}_{\T{1}\T{1}},\nn
&&\partial_\tau N^{\text{B}}_{\T{2}\T{2}} = -2\geta q^2\left(N^{\text{B}}_{\T{2}\T{2}}-\frac{Tw}{\tau}\right)-\frac{2(1+\sin^2\theta)}{\tau} N^{\text{B}}_{\T{2}\T{2}}.
\end{eqnarray}
The equations for $N^{\text{B}}_{\T{1}\T{1}}$ and
$N^{\text{B}}_{\T{2}\T{2}}$ match those in
Ref.~\cite{Martinez:2018}. The remaining equations, although very
similar, do not match completely. We believe our results are correct
but do not have a definitive explanation
for these disagreements.

Unlike the chargeless fluid in Ref.~\cite{An:2019rhf}, the charged fluid
in the present paper can be taken to nonrelativistic limit, where it
can be compared with
Ref.~\cite{Andreev:1978,andreev1980twoDliquids}.
The most glaring omission in
Ref.~\cite{Andreev:1978,andreev1980twoDliquids}
are the $G_{m \T{i}}$ components of the correlators.
It appears they were omitted based on the observation that their
equilibrium values vanish. They do, but they are not zero {\em out} of
equilibrium and are essential, for example, for describing the dominant
critical contribution to conductivity as discussed in
Section~\ref{sec:cond-its-freq}. 

To compare the equations for remaining components of $W_{AB}$, we need to
rescale our variables as
\begin{equation}
  N^{\text A}_{mm}\equiv\frac{W_{mm}}{n^2T^2}, \quad N^{\text A}_{\T{i}\T{j}}\equiv\frac{W_{\T{i}\T{j}}}{wn}.
\end{equation}
Omitting $W_{m \T{i}}$ terms as in Ref.~\cite{Andreev:1978,andreev1980twoDliquids}
we find
\begin{equation}\label{eq:Andreev}
\begin{aligned}
\mathcal L[N^{\text A}_{mm}]&=- 2\gl q^2\left(N^{\text A}_{mm}-\frac{c_p^{\text A}}{n}\right)+\theta N^{\text A}_{mm}, \\
\mathcal L[N^{\text A}_{\T{i}\T{j}}]&=- 2\gamma_\eta q^2\left(N^{\text A}_{\T{i}\T{j}}-\frac{T}{n}\delta_{ij}\right)+(1+c_s^2)\divu N^{\text A}_{\T{i}\T{j}}-(\theta^{\mu\nu}-\omega^{\mu\nu})\left(t_\mu^{(i)}t_\nu^{(k)}N^{\text A}_{\T{k}\T{j}}+t_\mu^{(j)}t_\nu^{(k)}N^{\text A}_{\T{i}\T{k}}\right),
\end{aligned}
\end{equation}
where the specific heat per mass is $c_p^{\text A}\equiv c_p/n$. This
 would agree nicely with Ref.~\cite{andreev1980twoDliquids}, if we
 also follow
 Ref.~\cite{andreev1980twoDliquids} and impose $N^{\text
  A}_{ij}\sim\delta_{ij}$ which will eliminate $\omega_{\mu\nu}$
term. Similar to the omission of  $G_{m \T{i}}$, the
assumption $G_{\T{i}\T{j}}\sim\delta_{ij}$ was apparently made by
neglecting off-equilibrium contribution to this correlator.

Our equation~\eqref{eq:Npm} for sound fluctuations
completely matches the Boltzmann equation given in
Ref.~\cite{Andreev:1978} in the nonrelativistic limit, where the
$\gp=\lambda c_s^2\alpha_p^2Tw$, appearing in Eq.~\eqref{eq:gammaL}
 is replaced by its
nonrelativistic limit
\begin{equation}
\gp^{\text{NR}}=\kappa\left(\frac{1}{c_v}-\frac{1}{c_p}\right).
\end{equation}
Indeed, since in our units the speed of light is 1, for a
non-relativistic fluid $c_s\ll1$ (see also
Table~\ref{tab:limits}).
\begin{equation}
  (Tn)^2\alpha_p^2=\left(1-\frac{\dot T}{c_s^2}\right)^2
\approx \frac{\dot T^2}{c_s^4}
=\frac{w}{c_s^2T}\left(\frac{1}{c_v}-\frac{1}{c_p}\right),
\end{equation}
where in writing `$\approx$' we used $c_s^2\ll\dot T$ (see
Table~\ref{tab:limits}) and expressed $\dot T$ in terms of $c_v$,
$c_p$ and $c_s$:
\begin{equation}
  \dot T^2=\frac{c_s^2w}{T}\left(\frac{1}{c_v}-\frac{1}{c_p}\right), \quad c_v\equiv\left(\frac{\partial\eps}{\partial T}\right)_n.
\end{equation}

\newcommand\Tstrut{\rule{0pt}{3.5ex}}         
\newcommand\Bstrut{\rule[-2ex]{0pt}{0pt}}   

    \begin{table}[ht]
    \centering
    \begin{tabularx}{1\textwidth}{>{\centering\arraybackslash}X |>{\centering\arraybackslash}X |>{\centering\arraybackslash}X >{\centering\arraybackslash}X >{\centering\arraybackslash}X }
    \hline \hline
            \parbox{7em}{thermodynamic quantities} & definition 
            & \parbox{7em}{conformal\\fluid}
            & \parbox{7em}{nonrelativistic\\ideal gas ($T\ll M$)}
            & \parbox{7em}{scaling\\power of $\xi$ at\\critical point}  \rule{0pt}{5ex}\rule[-3.6ex]{0pt}{0pt}\\
            \hline
            $c_s^2$ & $\left({\partial p}/{\partial\eps}\right)_m$
            & 1/3 & $\gamma T/M$ & $-\alpha/\nu$ \Tstrut\Bstrut\\ 
            \hline
            $c_p$ & $Tn\left({\partial m}/{\partial T}\right)_p$ &
            $c_p$ & ${\gamma n}/{(\gamma-1)}$ & $2-\eta$ \Tstrut\Bstrut\\
            \hline
            $\dot T$ & $\left({\partial\log T}/{\partial\log s}\right)_m$ & 1/3 & $\gamma-1$ & $-\alpha/\nu$ \Tstrut\Bstrut\\
            \hline
            $\dot c_s$ & $\left({\partial\log c_s}/{\partial\log s}\right)_m$ & 0 & $(\gamma-1)/2$ & $(1-\alpha)/\nu$ \Tstrut\Bstrut\\
            \hline
            $\dot c_p$ & $\left({\partial\log c_p}/{\partial\log s}\right)_m$ & 1 & $\gamma$ & $(1-\alpha)/\nu$  \Tstrut\Bstrut\\
    \hline \hline
    \end{tabularx}
        \caption{The behavior of thermodynamic coefficients used in
          the paper in different limits. For non-relativistic ideal
          gas $\gamma=c_p/c_v>1$ denotes adiabatic constant and $M$ the
          molecule mass.}\label{tab:limits}
    \end{table}


\bibliographystyle{utphys}
\bibliography{references}

\end{document}